\documentclass[%
 amsmath,amssymb,
 aps,
 prl,
noeprint,
floatfix,
reprint
]{revtex4-2}
\usepackage{graphicx,xcolor}
\definecolor{darkblue}{RGB}{0,0,150}
\definecolor{nightblue}{RGB}{0,0,100}

\usepackage{mathrsfs,dsfont,mathtools}

\usepackage{dcolumn}
\usepackage{bm}
\usepackage[
colorlinks,
citecolor=darkblue,
linkcolor=darkblue,
urlcolor=nightblue]{hyperref}

\usepackage[english]{babel}
\usepackage[babel,kerning=true,spacing=true]{microtype}

\usepackage{feynmp-auto}

\renewcommand{\Re}{\textrm{Re}}
\renewcommand{\Im}{\textrm{Im}}
\newcommand{\bk}{\mathbf{k}}

\begin{document}

\title{Nonvanishing sub-gap photocurrent
as a probe of lifetime effects}
\author{Daniel Kaplan}
\author{Tobias Holder}
\email{tobias.holder@weizmann.ac.il}
\author{Binghai Yan}
\affiliation{Department of Condensed Matter Physics,
Weizmann Institute of Science,
Rehovot 7610001, Israel}

\date{\today}

\begin{abstract}
For semiconductors and insulators, it is commonly believed that in-gap transitions into non-localized states are smoothly suppressed in the clean limit, i.e. at zero temperature their contribution vanishes due to the unavailability of states. 
We present a novel type of sub-gap response which shows that this intuition does not generalize beyond linear response. 
Namely, we find that the dc current due to the bulk photovoltaic effect can be finite and mostly temperature independent in an allowed window of sub-gap transitions. 
We expect that a moderate range of excitation energies lies between the bulk energy gap and the mobility edge where this effect is observable. 
Using a simplified relaxation time model for the band broadening, we find the  sub-gap dc-current to be temperature-independent for non-interacting systems but temperature-dependent for strongly interacting systems. Thus, the sub-gap response may be used to distinguish whether a state is single-particle localized or many-body localized.
\end{abstract}

\maketitle



\begin{figure}
    \centering
    \centering\includegraphics[width=.95\columnwidth]{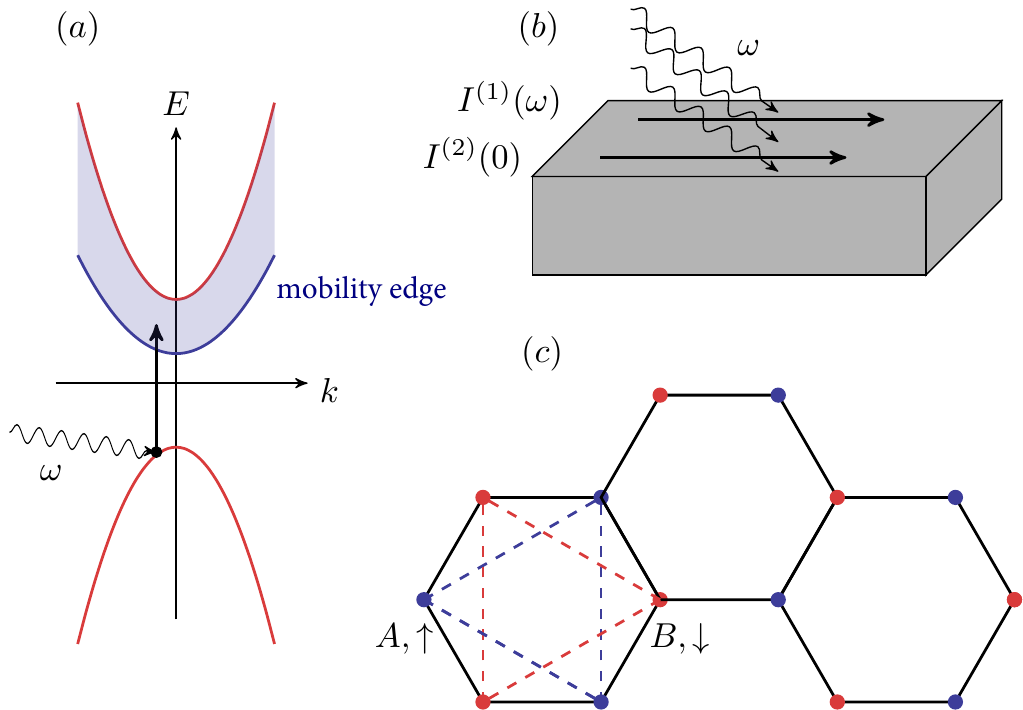}
    \caption{Main elements of the proposal. (a) Schematic of the processes leading to a photocurrent below the gap. The photon with sub-gap energy creates an excitation in the region between the mobility edge and the lower end of the conduction band. (b) Current response up to second order. The incident light of frequency $\omega$ creates a linear response current $I^{(1)}$ at frequency $\omega$. At 2\textsuperscript{nd} order, a photovoltaic current ($\omega=0$) is observed.
    (c) Sketch of the antiferromagnetic honeycomb lattice with next-nearest neighbor hoppings between sublattices $A$ and $B$ which is used as a model system.
    }
    \label{fig:scheme}
\end{figure}

\paragraph{Introduction.---}
Since its inception, it has been a fundamental tenet of the theory of optical response that an optical excitation requires both an occupied initial state from which to excite and an empty state into which to excite.
However, when going beyond linear optical response, it is possible for a quasiparticle to reoccupy a state that it vacated after a sequence of excitations into states of high energy~\cite{Sipe2000}. 
For example, under light irradiation a noncentrosymmetric solid usually exhibits two types of qualitatively different second-order responses, not only second harmonic frequency emission, but also bulk dc-current generation.
As we show in this letter, the latter process may involve off-resonant intermediate states, giving rise to a non-linear rectification current for frequencies which are too small to cross the gap between valence and conduction band. This results in a new type of non-linear dc-current response for certain magnetic materials, which would be optically transparent according to linear response theory (Fig.~\ref{fig:scheme}).

We propose to use this effect to investigate whether the band broadening in a given system is due to disorder or interactions. If additionally a mobility edge can be observed, non-linear optical response opens a new avenue to explore the physics of weak localization and also many-body localization~\cite{Alet2018,Abanin2019} in condensed matter systems.
This connection between optical response and localization effects is quite special, as we elucidate in the following.

Transport in non-interacting systems can be characterized in the language of Anderson localization by assigning a localization length to each state~\cite{Kramer1993,Evers2008}. 
The transition from infinite to finite localization length then immediately defines a mobility edge differentiating wavefunctions which do contribute to transport from those who do not. Most importantly, a decreasing density of states (DOS) is accompanied by a mobility edge separating delocalized states at high DOS from localized ones at low DOS~\cite{Mott1987}. 
However, for optical response of systems with a direct band gap $\Delta$, these localization effects are usually not important. 
Namely, upon irradiation with an electric field $\bm{E}(\omega)$ of frequency $\omega$, the ac-conductivity $\sigma^{aa}(\omega)$, associated with a current $j^a(\omega) = \sigma^{aa}(\omega)E_a(\omega)$ is given by the Kubo formula. Therefore, for $\omega<\Delta$, the conductivity $\sigma^{aa}(\omega)$ decreases smoothly as the density of states vanishes inside the bandgap with decreasing band broadening. 
This behavior is captured by assigning to eigenstates of energy $\varepsilon_m(\bm{k})$ with band index $m$ a finite linewidth $\tau_m$ through the so-called adiabatic switching $\omega\to\omega +i/2\tau_m$ in the inverse propagator $G_0^{-1}(\omega,\bm{k})=\omega+i0^+-\varepsilon_m(\bm{k})$. 
This band broadening captures well the diffusive nature of the quasiparticle motion, while it lacks more sophisticated corrections which go beyond self-energy effects, in particular it lacks localization corrections.
As it is well-known, at small temperatures, the band broadening will induce a finite DOS of size  $\mathcal{O}(1/\tau_m)$ in the gap of fully gapped insulators and semiconductors, which vanishes smoothly as $1/\tau_m\to 0$. The same holds true for the conductivity $\sigma^{aa}(\omega)$. For this reason, in the clean limit the current already vanishes even without localization, making the mobility edge very hard to resolve experimentally~\cite{Semeghini2015}.

In the theory of non-linear optical response current creation is likewise discussed without resorting to localization~\cite{Belinicher1980,vonBaltz1981,Sipe2000,Ventura2017,Parker2019,Holder2020}. 
It is a widely held preconception that one can smoothly recover a vanishing non-linear conductivity in insulators and semiconductors upon taking the clean limit, just like at linear order. Indeed, this has to be the case if all processes are adiabatic~\cite{Belinicher1986}. However, in the following we show that adiabaticity is not guaranteed for the second order rectification current in magnetic compounds, giving rise to a novel type of sub-gap response, which we argue to be a possible probe for the mobility edge.
In contrast to the continuous drive suggested here, pulsed optical probes will always lead to a sub-gap signal, but they present a drive which is intrinsically non-adiabatic, making an optical pulse insensitive to the lifetime effects reported here.

In the following, the current is driven by an extrinsic field-induced polarization as it typically arises in second-order optical response~\cite{
vonBaltz1981,Morimoto2016a,Tan2016,Cook2017,
Olbrich2009,Yuan2014,McIver2012,Okada2016,Koenig2017,Golub2018,deJuan2019,Bhalla2020}.
To this end, we study the dc-conductivity $\sigma^{aa;a}(0; \omega, -\omega)$ using the second-order perturbative expansion in the velocity gauge, defined through a dc-current density $j^c(0) = \sigma^{ab;c}(0; \omega, -\omega)E_a^{*} (-\omega)E_b(\omega)$ (cf. Fig.~\ref{fig:scheme}). 
We consider an insulator with a gap $\Delta$, whose states have a finite band broadening, parametrized by the lifetimes $\tau_m$ for each band. 
We further introduce the linewidth of intra- and interband processes, denoted $\gamma=1/\tau_v$ and $\Gamma=1/2\tau_v+1/2\tau_c$, with $\tau_{v,c}$ the lifetimes of valence and conduction band, respectively~\footnote{Unless stated otherwise, we do not specify whether these lifetimes are quasiparticle lifetimes or momentum lifetimes. Such distinction can be easily made from the diagrammatics~\cite{Holder2020}. Also note that the bulk-photovoltaic effect does not rely on carriers in the conduction band, which is why the interband scattering rate (recombination rate) does not enter.}. By expanding the expressions in the regime where $\omega < \Delta$, we find for systems with time-reversal symmetry (TRS) a suppression of the sub-gap conductivity in accordance with the expected scaling in the regime $\sigma^{aa;a}(\omega<\Delta) \sim \Gamma$, which vanishes smoothly in the clean limit. 
However, for systems lacking TRS the conductivity turns out to be non-vanishing as a function of $\Gamma,\gamma\to0$ where $\gamma/\Gamma$ fixed. 
As a  reason we note that the non-linear optical conductivity is the result of several different two- and three-band processes which normally interfere destructively, but which do no longer annihilate each other completely once TRS is broken. In optical response, such a interference between different processes is possible because the interaction with the electric field is, at least within some bounds, coherent~\cite{Holder2020}.
This sub-gap current can be viewed as off-resonant driving of virtual excitations at the band edge, which - at second order in the drive amplitude - does not necessarily vanish as fast as the DOS does.
Formulated in the language of second-order response theory, we observe that in the sub-gap regime the injection current equals the negative of the shift current only for TRS-preserving materials.
We emphasize that the absence of cancellation in the case of broken TRS represents a deviation from adiabatic driving, which is possible since the drive, even though a continuous wave, has finite frequency.

Since the conductivity in our result depends on the ratio  of intraband/interband relaxation rates $\gamma/\Gamma$, it is a sensitive measure of the interaction-induced modifications to the quasiparticle lifetimes. 
We believe that this sensitivity is specific enough to allow state-of-the-art experimental techniques to probe the physics of many-body localization in high-quality 2D and layered 3D materials.

\paragraph{Results.---}
The theory of second-order optical response is well established~\cite{Belinicher1980,vonBaltz1981,Sipe2000,Young2012}. Here, we employ a modern diagrammatic formulation which very transparently delineates intra- and interband lifetimes~\cite{Parker2019,Holder2020}.

For the electromagnetic interaction with the incident light we take the dipole approximation, which leads to three electron-photon vertices. For a Bloch Hamiltonian $H_0(\bm{k})$, they are the electron-photon vertex in direction $a$ with matrix elements $v_{nm}^a = \langle n|\partial_{k_a} H_0|m\rangle$ for Bloch band indices $m,n$, the two-electron-two-photon vertex $w^{ab}_{nm} = \langle n|\partial_{k_a} \partial_{k_b} H_0 |m\rangle$ and the five-leg vertex $u_{nm}^{abc} = \langle n| \partial_{k_a} \partial_{k_b} \partial_{k_c} H_0 | m\rangle$. The Fermi-Dirac factor is $f_m$, we further use shorthands $f_{mn}=f_m-f_n$ and $\varepsilon_{mn}=\varepsilon_m-\varepsilon_n$.

\begin{figure}
    \centering
    \includegraphics[width=\columnwidth]{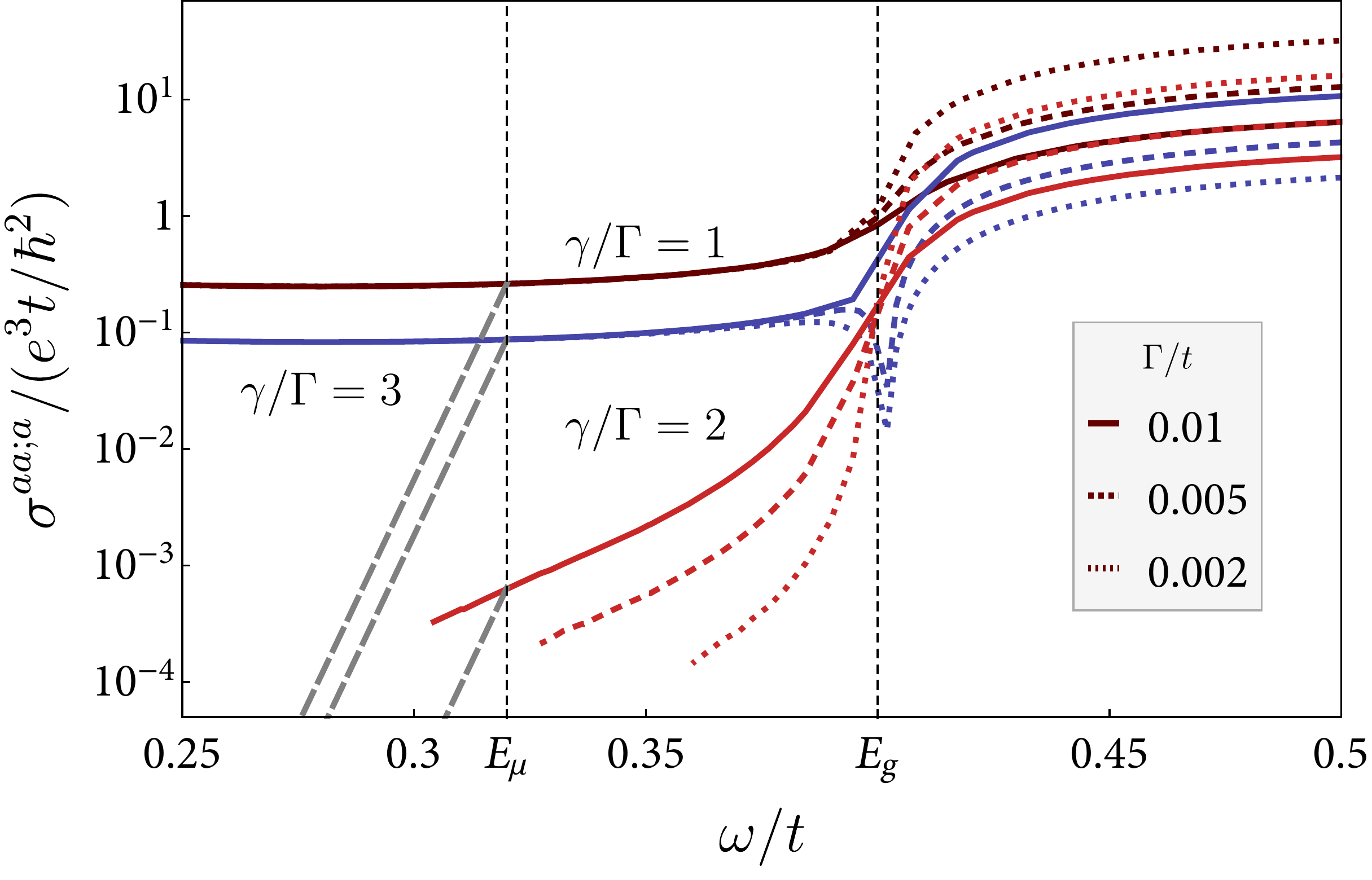}
    \caption{Modulus of the non-linear conductivity as a function of frequency for three different relaxation rates and three ratios $\gamma/\Gamma$ when TRS is broken. The gap is at $\Delta=0.4t$. Above the gap, the conductivity increases with increasing lifetime, below the gap the conductivity is generically independent of the lifetime and only depends on the ratio $\gamma/\Gamma$. The exception to this behavior is the fine-tuned value $\gamma/\Gamma=2$. 
    The gray dashed lines show schematically how taking localization physics into account will modify the result: It leads to an exponential decay of the current below the mobility edge $E_{\mu}$.
    The parameters used in the Hamiltonian Eq.~\eqref{eq:Ham} are $m_1=0.4t$, $m_2=0$ and $\delta=0.2t$.}
    \label{fig:1}
\end{figure}

\begin{figure}
    \centering
    \includegraphics[width=\columnwidth]{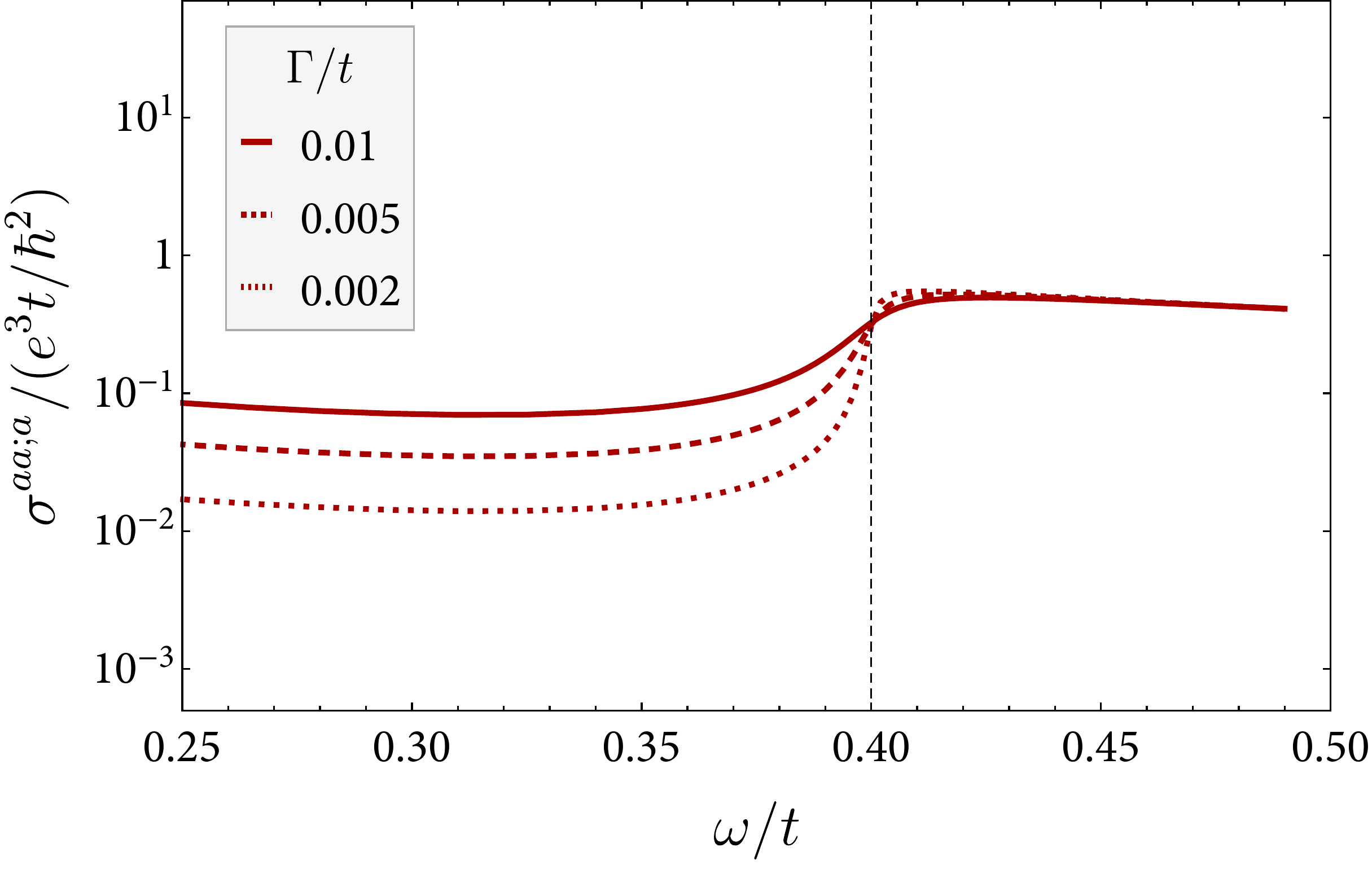}
    \caption{Modulus of the non-linear conductivity as a function of frequency for three different relaxation rates, but now for a system that preserves TRS. Above the gap, the conductivity is independent of the lifetime. It is always independent of the ratio $\gamma/\Gamma$. 
    The figure does not include any effects of localization.
    The parameters are chosen such that the gap is again at $\Delta=0.4t$. They are $m_1=0$, $m_2=0.4t$ and $\delta=0.2t$.}
    \label{fig:2}
\end{figure}

We assume in the following the existence of an energy gap $\Delta(\bm{k})$. 
For $\omega < \Delta$ we find that diagrams of the density terms cancel, $\int_\bk \sum_n f_n u_{nn}^{abc} = 
\int_\bk \sum_{nm}\frac{f_{nm}w^{ab}_{nm}v_{mn}^a}{\epsilon_{mn}}$~\footnote{See supplemental material for details of this derivation.}. 
Focusing on linear polarized light , we obtain two contributions $\eta_{1,2}$,
\begin{align}
    \eta_1 = 2 C \Re\Bigl[\frac{1}{i\gamma}\int_\bk \sum_{nm} f_{nm} \frac{|v_{nm}^a|^2v_{nn}^a}{\omega_1 -\epsilon_{mn}+i\Gamma} + (\omega_1 \to \omega_2)\Bigr],
\end{align}
where $C = - e^3/\omega^2\hbar^2$.
When TRS is present, $T\eta_1T^{-1} = -\eta_1$, and as expected this term vanishes identically. 
However, when $T$-symmetry is absent, $\eta_1$ can take a non-zero value.
The second contribution is derived from the vertex,
\begin{align}
    \eta_2 = C \Re\Bigl[\int_\bk\sum_{nm}f_{nm}\frac{w_{nm}^{aa}v_{mn}^a}{\omega_2 -\epsilon_{mn} + i\Gamma} + (\omega_2 \to \omega_1)\Bigr].
\end{align}
In what follows, we focus specifically on the limit $\omega_{1,2} \ll \Delta$, and use the replacement $v_{nm}^a = i \epsilon_{nm}r_{nm}^a$, where $n \neq m$. We first obtain that $\eta_1 = -4C \sum_{nm} f_{nm} \int_\bk \frac{\Gamma}{\gamma} |r_{nm}^a|^2v_{nn}^a$, after letting $\Gamma \to 0$. By using $w_{nm}^{aa} =\partial v^a_{nm} + i[v^a,r^a]_{nm}$, we find that the leading order component is then $\eta_2 = 2 C \sum_{nm} f_{nm} |r_{nm}^a|^2 v_{nn}^a$, where we considered only the real part of the numerator in expanding $\eta_2$ as the imaginary part is subleading. The total response below the gap, $\sigma_{sub}=\eta_1+\eta_2 $ therefore becomes
\begin{align}
    \sigma_{\mathrm{sub}}= -2C\left(\frac{2\Gamma}{\gamma}-1\right)\Biggl(\int_\bk\sum_{nm}f_{nm}|r_{nm}^a|^2v_{nn}^a\Biggr).
    \label{eq:main}
\end{align}
$\sigma_{\mathrm{sub}}$ is odd under TRS, and hence vanishes identically under an applied linearly polarized field. We therefore conclude that the T-symmetric system has a sub-gap dc-conductivity no larger than the next order terms, i.e. $\Gamma/\Delta$. 
Note that in this case, the smooth limit $\Gamma \to 0$ exists, meaning that $\sigma^{aa;a}\to 0$ independently of the precise finite value of the ratio $\gamma/\Gamma$. 
The main result of this letter is contained in Eq.~\eqref{eq:main}, which is the leading order term for a system lacking TRS. In the conventional language of non-linear optical response, Eq.~\eqref{eq:main} is the sum of the injection current ($\eta_1$) and the shift current ($\eta_2$). The factor of $\frac{2\Gamma}{\gamma}-1$ appearing in this relation implies that for the value of $\gamma/\Gamma=2$ the injection and shift currents cancel below the gap, up to regular terms. 
For any other value of $\gamma/\Gamma$ it remains finite. 
Since the dc-current is a function of the dimensionless parameter $\gamma/\Gamma$, it is insensitive to the absolute values of the broadening induced by $\gamma, \Gamma$. Most notably, the clean limit of $\Gamma,\gamma \to 0$ with $\gamma/\Gamma$ fixed does not lead to a smoothly vanishing dc-current according to Eq.~\eqref{eq:main}. 
By no means this implies that the current persists at arbitrarily small frequencies below the gap, as these considerations do not consider
the effects of localization. Yet, the result does not require a specific microscopic model for the type of broadening which leads to the finite relaxation rates $\gamma, \Gamma$, except to require that these rates can be approximated by some relaxation time approximation~\footnote{In materials where the transport time and quasiparticle lifetime are dissimilar, it is necessary to also account for possible vertex corrections. This does not change the main observation put forth here that only the adiabatic limit leads to a vanishing sub-gap response.}. We propose to make use of this desirable property as a probe of localization physics.
\paragraph{Example.---}
For demonstration, we consider a 2d Haldane model with next-nearest neighbor spin-orbit-like coupling, similar to the form suggested by Kane and Mele~\cite{Kane2005,Bernevig2006}. 
In an infinite 2d-system almost all eigenstates are localized, meaning that there is no mobility edge. However, it is still sensible to discuss currents and localization effects for mesoscopic systems with a size smaller than the largest localization length~\cite{Licciardello1975}. 
Our Hamiltonian reads,
\begin{align}
    H = t_1 \sigma_x + t_2 \sigma_y + m_1 \sigma_z s_z + m_2 \sigma_z + \delta(k) \sigma_z s_z.
    \label{eq:Ham}
\end{align}
Where $t_1 = \Re[t(k)], t_2 = \Im[t(k)]$, such that \begin{align*}
t(k)=  -t\left(e^{i\left(\frac{3}{2}k_x+k_y\right)}+e^{i\left(-\frac{3}{2}k_x+k_y\right)}+e^{-ik_y}\right).
\end{align*} 
The nearest-neighbor distance of the honeycomb lattice is $b = 1$, $\sigma$ and $s$ relate to the sub-lattice and spin degrees of freedom, respectively. Inversion in this system takes the form $P = \sigma_x K$, where $K: \bm{k} \to -\bm{k}$, and time-reversal is $T =i s_y \mathcal{C} K$, $\mathcal{C}$ corresponds to complex conjugation.  We set $m_1 \neq 0, ~ m_2 = 0$ for TRS-breaking, and $m_1 = 0, ~ m_2 \neq 0$ for the T-symmetric case. The spin-orbit coupling term has the form,
\begin{align}
\delta(k) &= -t \delta \Bigl[ \sin\left(\sqrt{3}k_x\right)  -\sin\left(\sqrt{3}k_x/2-3k_y/2\right)- 
\notag\\&\qquad
\sin\left(\sqrt{3}k_x/2+3k_y/2\right)\Bigr].
\end{align}
Fig.~\ref{fig:1} shows the results for the conductivity $\sigma^{xx;x}=-\sigma^{xy;y}$, for linearly polarized light using $m_1 \neq 0$. Crystal symmetries dictate that $\sigma_{xx;x} = -\sigma^{yy;x}=-\sigma^{yx;y}$, generally, while $\sigma^{yy;y} =0$, and hence $\sigma^{xx;y} = \sigma^{yx;x}=\sigma^{xy;x} =0$. Several different ratios $\gamma/\Gamma$ are shown, with the gap presented as the dashed line. We find that for any ratio except $\gamma/\Gamma = 2$, the conductivity below the gap quickly converges to a non-zero value that depends on the ratio $\gamma/\Gamma$. In particular, for values $\gamma/\Gamma>2$, we find that the conductivity changes sign as the frequency crosses the gap. It is only for the ratio $\gamma/\Gamma = 2$ that we observe a scaling as $\sim \Gamma^2$, in agreement with Eq. \eqref{eq:main}, it is this ratio that the leading order term $\eta_1+\eta_2$ disappears and only sub-leading terms remain. 
In conclusion, the subgap response is markedly different between $\gamma/\Gamma = 2$ and $\gamma/\Gamma \neq 2$, with asymptotics $\sigma^{aa;a}\sim \mathcal{O}(\omega^{-2})$ and $\sigma^{aa;a}\sim \mathcal{O}(\omega^{-2}\Gamma^2)$ respectively.
In Fig.~\ref{fig:2}, the conductivity component $\sigma^{yy;y}=-\sigma^{xx;y}$ is shown for $m_2 \neq 0$, which is the only one present due to mirror symmetries and TRS.
Due to TRS, we recover the expected result that $\sigma \sim \Gamma$ below the gap. Notably, the conductivity can smoothly be continued to $\Gamma \to 0$, and the result is independent of the ratio $\gamma/\Gamma$. For values above the gap, Fig.~\ref{fig:1} shows the appearance of the injection current, which scales as $\frac{1}{\gamma}$, while in Fig.~\ref{fig:2} we see the shift current emerging, which for $\Delta \ll \omega$ is independent of $\Gamma$. 
We also checked numerically that $\eta_1 + \eta_2$ compares well to the complete second order response as a function of the ratio $\gamma/\Gamma$, which confirms that the sub-gap conductivity is dominated by this term.
As has been noted in~\cite{Passos2018}, the ratio of $\gamma/\Gamma = 2$ is the result of assuming a particular relaxation mechanism for the different bands. This condition corresponds to the replacement in the response of $\omega_i \to \omega_i + i\Gamma$, such that in  denominators with two frequencies one should put $\omega_1+\omega_2 \to \omega_1+\omega_2 + 2i\Gamma$.
However, this prescription is not always justified. 
Namely, the ratio $\gamma/\Gamma$ is
\begin{align}
\frac{\gamma}{\Gamma}&= \frac{\frac{1}{\tau_v}}{\frac{1}{2\tau_v}+\frac{1}{2\tau_c}}. 
\end{align}
In order to fulfill the condition $\gamma/\Gamma \sim 2$ which enforces the cancellation in the TRS-broken case, it must thus hold that $\tau_c/\tau_v \to \infty$. 
In other words, the quasiparticle in the empty conduction band has a lifetime greatly exceeding the ground state quasiparticle lifetime. There is essentially only one situation in which this seems reasonable, which is for a strongly-correlated ground state.

We note that ratios such that $\gamma/\Gamma > 2$ are even harder to achieve since this would imply that the coherence time of the interband transition is significantly longer than that of the intraband one. 
This phenomenology also implies that the current originating from $\sigma_{sub}$ is directly related to the departure from adiabaticity, and in fact quantifies the degree to which the adiabatic approximation with the successive limits $\tau_1\to\infty$ and then $\tau_0\to\infty$ fails to hold.

For concreteness, we now use a two-parameter form for the temperature dependence of the relaxation rates, i.e. $\gamma = c_{00}(T) + c_{01}T^{n}$,  and $\Gamma = c_{10}(T)+c_{11} T^{n}$. 
Here, the first part is due to static disorder or electron-phonon interactions, while the $T^n$ piece is due to electron-electron interactions.
In the usual fashion, the factors $c_{i0}(0)$ describe the residual scattering induced by static disorder in the system. 
At low temperatures, the ratio approaches $\gamma/\Gamma = c_{10}(T)/c_{00}(T)\sim 1$, as the elastic mean free path due to disorder and phonons is insensitive to the occupation numbers in valence and conduction band.
However, for a strongly correlated system with low-disorder this will cross over into $\gamma/\Gamma = c_{11}/c_{01}\sim 2$ since the inelastic electron-electron interaction is proportional to the occupied DOS in the densely populated valence band, enforcing $\tau_v/\tau_c\to 0$. 
In both cases, the sub-gap conductivity is  eventually intercepted by localization effects as a function of decreasing frequency.
However, the onset of such activated behavior is markedly dissimilar. 
In the interacting case, decreasing the temperature will decrease both $\gamma$ and $\Gamma$ so that $\gamma/\Gamma =2$, therefore the current decreases for any frequency below the gap (cf. Fig~\ref{fig:1}). In contradistinction, in the non-interacting case a decrease in temperature will again decrease both $\gamma$ and $\Gamma$ but with $\gamma/\Gamma \sim 1$. Then, the conductivity at frequencies between the mobility edge and the gap will remain unchanged.
While there is limited temperature window in a strongly interacting system where the lifetime of states in the conduction band are limited by electron-electron interactions, i.e. where $c_{10}(T)<c_{11}T^n$, the phenomenology presented here always allows to directly identify whether such a window exists from the temperature dependence of the sub-gap current.
One might use the same reasoning to determine whether or not the mobility edge - if it is observable - appears for states which are well described in a single-article picture or rather ones which follow from a strongly coupled many-body description.
Also here, there are some restrictions on the suitable temperature range and the inelastic mean free path as the mobility edge is usually located a distance $V_0^2/\Lambda$ from the corresponding band edge, with $V_0$ the disorder strength and $\Lambda$ the bandwidth~\cite{Piraud2013,Semeghini2015,Mueller2016}.
Of course, when TRS is present it instead holds true that $\sigma \sim \Gamma$, resulting in a conductivity which carries no information about the ratio $\gamma/\Gamma$. 
In this latter case, one cannot separate the different sources of relaxation without exploring the microscopics of the electron motion.

In the discussion we disregarded further corrections other than self-energy terms. While their inclusion may alter the parametric value of the lifetimes entering into $\gamma$ and $\Gamma$~\cite{Mahan1990}, it does not affect our main observation that only a special ratio of intra- and interband relaxation rates restores the result expected from adiabaticity.

\paragraph{Conclusions.---}
In this letter, we investigated the second-order dc-conductivity in response to irradiation with light of frequency below the energy gap of a semiconductor or insulator lacking time-reversal symmetry. Our results show that the magnitude and frequency dependence of the conductivity is qualitatively different from the time-reversal symmetric case. Using a simple model for the quasiparticle relaxation, we proposed a way to probe the  lifetimes in experimentally realizable high-quality samples. 
Namely, we predict a large and temperature independent sub-gap dc-current for non-interacting quasiparticles and a T-dependent suppression of the bulk photovoltaic response for strong interactions. 
If a sub-gap mobility edge is observed in this latter case, it can be associated with localization originating from a many-body state.

The bulk photovoltaic effect has been studied only for few magnetic compounds~\cite{Ganichev2002,Ganichev2006,Zhang2019,Sun2019}.
Good material candidates to observe the proposed effect are transition metal dichalcogenide monolayers of $\mathrm{MnPX_{3}}$, $\mathrm{X=Se,\, S}$~\cite{Li2012} and some hexagonal Mott insulators like $\mathrm{RuCl_3}$~\cite{Koitzsch2016,Zhou2016}. Also, by measuring the second order conductivity above and below the Néel temperature, it is possible to explore the response with and without TRS in the same sample.

\begin{acknowledgments}
We thank 
D. E. Parker, 
D. Passos, and
A. Stern
for helpful discussion. 
B.Y. acknowledges the financial support by the Willner Family Leadership Institute for the Weizmann Institute of Science, the Benoziyo Endowment Fund for the Advancement of Science,  Ruth and Herman Albert Scholars Program for New Scientists, the European Research Council (ERC) under the European Union's Horizon 2020 research and innovation programme (Grant No. 815869).
\end{acknowledgments}

%

\newpage
\pagebreak
\onecolumngrid
\newpage
\begin{center}
\textbf{\large Supplemental Material: Nonvanishing sub-gap photocurrent
as\protect\\ a probe of lifetime effects}
\end{center}
\setcounter{equation}{0}
\setcounter{figure}{0}
\setcounter{table}{0}
\makeatletter
\renewcommand{\theequation}{S\arabic{equation}}
\renewcommand{\thefigure}{S\arabic{figure}}

\section{Complete second-order conductivity}
The results in the main text quoting $\sigma^{aa;a}$ are derived from results obtained in Ref. \cite{Holder2020,Parker2019}. 
For the derivation consider the following perturbative process: the Bloch Hamiltonian $H_0 (\mathbf{k})$ is perturbed using the minimal coupling scheme, via $\mathbf{k} \to \mathbf{k} + e\mathbf{A}$, where $A$ is the vector potential. The resulting Hamiltonian is then expanded in small $\mathbf{A}$ as, $H(\mathbf{k}) = H_0(\mathbf{k}) + \partial_a H_0 (\mathbf{k}) A_a + \frac{1}{2} \partial_a \partial_b H_0 (\mathbf{k}) A_a A_b + \ldots$, where $\partial_a = \frac{\partial}{\partial k_a}$. At second order, this expansion amounts to an quasiparticle loop with two incoming photons (at frequencies $\omega_1, \omega_2$), and with a current vertex at frequency $\bar \omega = \omega_1+\omega_2$. A detailed overview of the diagrammatics is provided in \cite{Parker2019,Holder2020}. 
The matrix elements of the vertices are defined as $v_{nm}^a = \langle n|\partial_a H_0|m\rangle$, $w^{ab}_{nm} = \langle n|\partial_a \partial_b H_0 |m\rangle$, and finally $u_{nm}^{abc} = \langle n| \partial_a \partial_b \partial_c H_0 | m\rangle$. $f_n$ is the Fermi factor, and the notation $f_{nm} = f_n - f_m$. $\varepsilon_n$ is the energy of the n-th Bloch band, and $\varepsilon_{nm} = \varepsilon_n - \varepsilon_m$.
In the most general case, $\sigma^{ab;c}$ reads, 
\begin{align}
    \sigma^{ab;c}&(\bar\omega,\omega_1,\omega_2)
    \notag\\&=
    \frac{-e^3}{\hbar^2 \omega_1\omega_2}
    \sum_{m,n,l}\int_{\bm{k}}
   f_m u_{mm}^{abc}
    +f_{mn} \frac{v_{mn}^a w_{nm}^{cb}}{\omega_1+\varepsilon_{mn}}
    +f_{mn} \frac{v_{mn}^b w_{nm}^{ca}}{\omega_2+\varepsilon_{mn}}
    +f_{mn} \frac{w_{mn}^{ab} v_{nm}^c}{\bar\omega+\varepsilon_{mn}}
    \notag\\&
    +\biggl(\frac{f_{mn}  v_{mn}^a v_{nl}^b v_{lm}^c}
    {(\omega_1-\varepsilon_{nm})(\bar\omega-\varepsilon_{lm}
    )}
    +\frac{f_{mn}  v_{ln}^a v_{nm}^b v_{ml}^c}
    {(\omega_2-\varepsilon_{mn})(\bar\omega-\varepsilon_{ml}
    )}+(a,\omega_1\leftrightarrow b,\omega_2)\biggr),
    \label{eq:generalseco}
\end{align}
for a current in the $c$ direction, with an applied electric field in the $a,b$ directions. 

\section{Immediate cancellations}
As stated in the main text, we place the chemical potential within the gap of an insulator. Therefore, at $T=0$, $f_n = 1$, if $\varepsilon_n - \mu < 0$ and $f_n =0$ otherwise. Consequently, we can directly calculate $\int_{\bf{k}} u_{nm}^{abc}$. Using the generalized derivative, $f_n u^{abc}_{nn} = f_n\partial_c w^{ab}_{nn} + if_n [w^{ab}, r^c]_{nn}$. The first term vanishes as it is a total derivative over the Brillouin zone (whenever the chemical potential is within the gap). The latter is resolved, leading to $f_n [w^{ab}, r^c]_{nn} = f_n \left(w^{ab}_{nm}r^c_{mn} - r^c_{nm}w^{ab}_{mn}\right)$. For the second term, we replace $m \leftrightarrow n$, and have, $f_n [w^{ab}, r^c]_{nn} = f_n w^{ab}_{nm}r^c_{mn} - f_m w^{ab}_{nm}r^c_{mn} = f_{nm}w^{ab}_{nm}r^c_{mn}$. Finally, using $v^c_{nm} = ir_{nm}^c\varepsilon_{nm}$, we arrive at $\int_{\bf{k}} u_{nm}^{abc} = \int_{\bf{k}} f_{nm}w_{nm}^{ab}v^c_{mn} \varepsilon_{mn}^{-1}$. This expression is exactly identical to the fourth term in Eq. \ref{eq:generalseco},  whenever $\bar \omega = 0$, albeit with a minus sign, thus cancelling it. 
\section{Derviation of $\eta_1, \eta_2$}
We derive the following  explicitly for the conductivity under a linearly-polarized electric field, in the $a$ direction; this is denoted by $\sigma^{aa;a}$. 
The DC response is given by $\omega_1 = -\omega_2 = \omega$, and $\bar \omega = 0$. As explained in the main text, we insert finite lifetimes according to the prescription $\bar \Omega \to \bar \Omega + i\gamma$, and $\omega_{1,2} -\varepsilon_{nm} \to \omega_{1,2}-\varepsilon_{nm}  + i\Gamma$.
We first tackle the terms containing a production of three $v^a_{nm}$ vertices. This is given by 
\begin{align}
    C \int_\bk \sum_{n,m,l} \biggl(\frac{f_{mn}  v_{mn}^a v_{nl}^a v_{lm}^a}
    {(\omega_1-\varepsilon_{nm})(\bar\omega-\varepsilon_{lm}
    )}
    +\frac{f_{mn}  v_{ln}^a v_{nm}^a v_{ml}^a}
    {(\omega_2-\varepsilon_{mn})(\bar\omega-\varepsilon_{ml}
    )}+(a,\omega_1\leftrightarrow b,\omega_2)\biggr).
    \label{eq:trip_v}
\end{align}
We then isolate the leading order component in $1/\gamma$ from Eq. \ref{eq:trip_v}. The leading order is obtained whenever $\varepsilon_{lm} =0$. Thus, with $C = -\frac{e^3}{\hbar^2\omega^2}$,
\begin{align}
    \eta_1 = C\Re\left( \sum_{nm}\int_{\bf{k}}f_{nm}\left(\frac{v^a_{nm}v^a_{mn}v^a_{nn}}{\left(\bar \omega + i\gamma\right)\left(\omega_1-\varepsilon_{nm}+i\Gamma\right)} + \frac{v^a_{nm}v^a_{mn}v^a_{mm}}{\left(\bar \omega + i\gamma\right)\left(\omega_2-\varepsilon_{mn}+i\Gamma\right)} \right)+ (\omega_1 \to \omega_2)\right).
\end{align}
After exchanging indices twice, and explicitly substituting $\bar \omega = 0$ for the DC response, we find Eq. 1 in the main text,
\begin{align}
    \eta_1 = -4C\Re\left( \int_{\bf{k}}\sum_{nm}f_{nm}\frac{|v_{nm}^a|^2 v_{nn}}{i\gamma\left(\omega_1 - \varepsilon_{nm} + i\Gamma\right)}\right).
    \label{eq:supeta1}
\end{align}
The case of $m \neq l$, is dealt with separately, below.
For the remaining contributions, we first consider the two terms in the first line of Eq.~\eqref{eq:generalseco}, which both contain the vertex $w$.
In the limit $\omega_1 \ll \Delta$, we expand the denominator in powers of $\omega/\Delta$. To order $\mathcal{O}\left(\frac{\omega}{\Delta}\right)$, we have $\int_\bk \sum_{nm} \varepsilon_{mn}^{-1} f_{nm} w^{aa}_{nm}v^{a}_{mn}$. 
We note that due to the difference of Fermi factors $f_{nm}$ only $n\neq m$ is nonzero. Using the generalized metric connection, $w^{aa}_{nm} = \partial_a v_{nm}^a + i[v^a,r^a]_{nm}$. $v_{nm}^a$ is once again amenable to the replacement $v^a_{mn} = i\varepsilon_{mn} r^a_{mn}$. This leaves
\begin{align}
    -\int_{\bk} \sum_{nm} f_{nm}\Delta_{nm}^a|r_{nm}^a|^2- \int_{\bk} \sum_{n,m} f_{nm}\varepsilon_{nm}\partial_a r_{nm}^a r^a_{mn}-\int_{\bk} \sum_{nm} f_{nm}[v^a,r^a]_{nm}r_{mn}^a,
    \label{eq:finalterm}
\end{align}
with $\Delta^a_{nm} = v^a_{nn} - v^a_{mm} = \partial_a \varepsilon_{nm}$.
The second term is written using a symmetrization $\int_{\bk} \sum_{n,m} f_{nm}\varepsilon_{nm}\partial_a r^a_{nm} r^a_{mn} = \frac{1}{2}\int_{\bk} \sum_{nm} f_{nm}\varepsilon_{nm}(\partial_a r_{nm}^a r^a_{mn} + r^a_{nm}\partial_a r_{mn}^a) =  \frac{1}{2}\int_{\bk} \sum_{nm} f_{nm}\varepsilon_{nm}(\partial_a |r^a_{nm}|^2)$. After integration by parts, we obtain the first term of Eq. \ref{eq:finalterm}, with a factor of $\frac{1}{2}$ and the opposite sign (note that $\partial_a f_{nm} = 0)$. Finally, the third term in Eq. \ref{eq:finalterm} has the form $\sum_{n,m,l} f_{nm}r_{mn}^a\left(v^a_{nl}r^a_{lm} - r^a_{nl}v^a_{lm}\right) =\sum_{n,m,l} f_{nm}\left(v^a_{nl}r^a_{lm}r_{mn}^a + r^a_{ml}v^a_{ln}r^a_{nm}\right)$. We separate the latter sum into two cases. When $l=n$, we obtain $2\int_{\bk}\sum_{n,m}f_{nm} |r^a_{mn}|^2 v^a_{nn} = \int_{\bk}\sum_{n,m}f_{nm} \Delta_{nm}^a|r^a_{nm}|^2$, contributing to the first term in Eq. \ref{eq:finalterm}. When $l \neq n$, we have $\int_{\bk}\sum_{n \neq m, l\neq m}f_{nm}i\epsilon_{nl}(r^a_{nl}r^a_{lm}r^a_{mn}-r^a_{ml}r^a_{ln}r^a_{nm})$. This is in turn directly cancelled by the $m \neq l$ found in Eq. \ref{eq:trip_v}. Since $m \neq l, n \neq m$, we are free to replace in Eq. \ref{eq:trip_v} $v^a_{nm} = i \epsilon_{nm} r_{nm}^a$ everywhere. Consequently, we have, $\int_\bk \sum_{n,m,l} i f_{mn} \epsilon_{nl} \left( r^a_{mn}r_{nl}^a r^a_{lm} - r^a_{nm}r^a_{ln}r^a_{ml}\right)$, which precisely cancels the preceding term.
Lastly, in order to conform to Eq.~\eqref{eq:supeta1}, we note that $\sum_{n,m}f_{nm}\Delta_{nm}^a |r_{nm}^a|^2 = 2 \sum_{n,m}f_{nm}v_{nn}^a |r_{nm}^a|^2$.
And therefore, the addition of all terms leads to
\begin{align}
    \eta_2 = C \Re\Bigl[\int_\bk\sum_{nm}f_{nm}\frac{w_{nm}^{aa}v_{mn}^a}{\omega_2 -\epsilon_{mn} + i\Gamma} + (\omega_2 \to \omega_1)\Bigr] \approx 2C \int_{\bf{k}} \sum_{nm} f_{nm} |r_{nm}^a|^2 v_{nn}^a,
\end{align}
which together with $\eta_1$, yields Eq.~(3) of the main text.

\begin{thebibliography}{42}%
\makeatletter
\providecommand \@ifxundefined [1]{%
 \@ifx{#1\undefined}
}%
\providecommand \@ifnum [1]{%
 \ifnum #1\expandafter \@firstoftwo
 \else \expandafter \@secondoftwo
 \fi
}%
\providecommand \@ifx [1]{%
 \ifx #1\expandafter \@firstoftwo
 \else \expandafter \@secondoftwo
 \fi
}%
\providecommand \natexlab [1]{#1}%
\providecommand \enquote  [1]{``#1''}%
\providecommand \bibnamefont  [1]{#1}%
\providecommand \bibfnamefont [1]{#1}%
\providecommand \citenamefont [1]{#1}%
\providecommand \href@noop [0]{\@secondoftwo}%
\providecommand \href [0]{\begingroup \@sanitize@url \@href}%
\providecommand \@href[1]{\@@startlink{#1}\@@href}%
\providecommand \@@href[1]{\endgroup#1\@@endlink}%
\providecommand \@sanitize@url [0]{\catcode `\\12\catcode `\$12\catcode
  `\&12\catcode `\#12\catcode `\^12\catcode `\_12\catcode `\%12\relax}%
\providecommand \@@startlink[1]{}%
\providecommand \@@endlink[0]{}%
\providecommand \url  [0]{\begingroup\@sanitize@url \@url }%
\providecommand \@url [1]{\endgroup\@href {#1}{\urlprefix }}%
\providecommand \urlprefix  [0]{URL }%
\providecommand \Eprint [0]{\href }%
\providecommand \doibase [0]{https://doi.org/}%
\providecommand \selectlanguage [0]{\@gobble}%
\providecommand \bibinfo  [0]{\@secondoftwo}%
\providecommand \bibfield  [0]{\@secondoftwo}%
\providecommand \translation [1]{[#1]}%
\providecommand \BibitemOpen [0]{}%
\providecommand \bibitemStop [0]{}%
\providecommand \bibitemNoStop [0]{.\EOS\space}%
\providecommand \EOS [0]{\spacefactor3000\relax}%
\providecommand \BibitemShut  [1]{\csname bibitem#1\endcsname}%
\let\auto@bib@innerbib\@empty
\bibitem [{\citenamefont {{Sipe}}\ and\ \citenamefont
  {{Shkrebtii}}(2000)}]{Sipe2000}%
  \BibitemOpen
  \bibfield  {author} {\bibinfo {author} {\bibfnamefont {J.~E.}\ \bibnamefont
  {{Sipe}}}\ and\ \bibinfo {author} {\bibfnamefont {A.~I.}\ \bibnamefont
  {{Shkrebtii}}},\ }\bibfield  {title} {\bibinfo {title} {{Second-order optical
  response in semiconductors}},\ }\href
  {https://doi.org/10.1103/PhysRevB.61.5337} {\bibfield  {journal} {\bibinfo
  {journal} {Phys. Rev. B}\ }\textbf {\bibinfo {volume} {61}},\ \bibinfo
  {pages} {5337} (\bibinfo {year} {2000})}\BibitemShut {NoStop}%
\bibitem [{\citenamefont {{Alet}}\ and\ \citenamefont
  {{Laflorencie}}(2018)}]{Alet2018}%
  \BibitemOpen
  \bibfield  {author} {\bibinfo {author} {\bibfnamefont {F.}~\bibnamefont
  {{Alet}}}\ and\ \bibinfo {author} {\bibfnamefont {N.}~\bibnamefont
  {{Laflorencie}}},\ }\bibfield  {title} {\bibinfo {title} {{Many-body
  localization: An introduction and selected topics}},\ }\href
  {https://doi.org/10.1016/j.crhy.2018.03.003} {\bibfield  {journal} {\bibinfo
  {journal} {CR Phys.}\ }\textbf {\bibinfo {volume} {19}},\ \bibinfo {pages}
  {498} (\bibinfo {year} {2018})}\BibitemShut {NoStop}%
\bibitem [{\citenamefont {{Abanin}}\ \emph {et~al.}(2019)\citenamefont
  {{Abanin}}, \citenamefont {{Altman}}, \citenamefont {{Bloch}},\ and\
  \citenamefont {{Serbyn}}}]{Abanin2019}%
  \BibitemOpen
  \bibfield  {author} {\bibinfo {author} {\bibfnamefont {D.~A.}\ \bibnamefont
  {{Abanin}}}, \bibinfo {author} {\bibfnamefont {E.}~\bibnamefont {{Altman}}},
  \bibinfo {author} {\bibfnamefont {I.}~\bibnamefont {{Bloch}}},\ and\ \bibinfo
  {author} {\bibfnamefont {M.}~\bibnamefont {{Serbyn}}},\ }\bibfield  {title}
  {\bibinfo {title} {{Colloquium: Many-body localization, thermalization, and
  entanglement}},\ }\href {https://doi.org/10.1103/RevModPhys.91.021001}
  {\bibfield  {journal} {\bibinfo  {journal} {Rev. Mod. Phys.}\ }\textbf
  {\bibinfo {volume} {91}},\ \bibinfo {eid} {021001} (\bibinfo {year}
  {2019})}\BibitemShut {NoStop}%
\bibitem [{\citenamefont {{Kramer}}\ and\ \citenamefont
  {{MacKinnon}}(1993)}]{Kramer1993}%
  \BibitemOpen
  \bibfield  {author} {\bibinfo {author} {\bibfnamefont {B.}~\bibnamefont
  {{Kramer}}}\ and\ \bibinfo {author} {\bibfnamefont {A.}~\bibnamefont
  {{MacKinnon}}},\ }\bibfield  {title} {\bibinfo {title} {{Localization: theory
  and experiment}},\ }\href {https://doi.org/10.1088/0034-4885/56/12/001}
  {\bibfield  {journal} {\bibinfo  {journal} {Rep. Prog. Phys.}\ }\textbf
  {\bibinfo {volume} {56}},\ \bibinfo {pages} {1469} (\bibinfo {year}
  {1993})}\BibitemShut {NoStop}%
\bibitem [{\citenamefont {{Evers}}\ and\ \citenamefont
  {{Mirlin}}(2008)}]{Evers2008}%
  \BibitemOpen
  \bibfield  {author} {\bibinfo {author} {\bibfnamefont {F.}~\bibnamefont
  {{Evers}}}\ and\ \bibinfo {author} {\bibfnamefont {A.~D.}\ \bibnamefont
  {{Mirlin}}},\ }\bibfield  {title} {\bibinfo {title} {{Anderson
  transitions}},\ }\href {https://doi.org/10.1103/RevModPhys.80.1355}
  {\bibfield  {journal} {\bibinfo  {journal} {Rev. Mod. Phys.}\ }\textbf
  {\bibinfo {volume} {80}},\ \bibinfo {pages} {1355} (\bibinfo {year}
  {2008})}\BibitemShut {NoStop}%
\bibitem [{\citenamefont {{Mott}}(1987)}]{Mott1987}%
  \BibitemOpen
  \bibfield  {author} {\bibinfo {author} {\bibfnamefont {N.}~\bibnamefont
  {{Mott}}},\ }\bibfield  {title} {\bibinfo {title} {{The mobility edge since
  1967}},\ }\href {https://doi.org/10.1088/0022-3719/20/21/008} {\bibfield
  {journal} {\bibinfo  {journal} {J. Phys. C Solid State Phys.}\ }\textbf
  {\bibinfo {volume} {20}},\ \bibinfo {pages} {3075} (\bibinfo {year}
  {1987})}\BibitemShut {NoStop}%
\bibitem [{\citenamefont {{Semeghini}}\ \emph {et~al.}(2015)\citenamefont
  {{Semeghini}}, \citenamefont {{Landini}}, \citenamefont {{Castilho}},
  \citenamefont {{Roy}}, \citenamefont {{Spagnolli}}, \citenamefont
  {{Trenkwalder}}, \citenamefont {{Fattori}}, \citenamefont {{Inguscio}},\ and\
  \citenamefont {{Modugno}}}]{Semeghini2015}%
  \BibitemOpen
  \bibfield  {author} {\bibinfo {author} {\bibfnamefont {G.}~\bibnamefont
  {{Semeghini}}}, \bibinfo {author} {\bibfnamefont {M.}~\bibnamefont
  {{Landini}}}, \bibinfo {author} {\bibfnamefont {P.}~\bibnamefont
  {{Castilho}}}, \bibinfo {author} {\bibfnamefont {S.}~\bibnamefont {{Roy}}},
  \bibinfo {author} {\bibfnamefont {G.}~\bibnamefont {{Spagnolli}}}, \bibinfo
  {author} {\bibfnamefont {A.}~\bibnamefont {{Trenkwalder}}}, \bibinfo {author}
  {\bibfnamefont {M.}~\bibnamefont {{Fattori}}}, \bibinfo {author}
  {\bibfnamefont {M.}~\bibnamefont {{Inguscio}}},\ and\ \bibinfo {author}
  {\bibfnamefont {G.}~\bibnamefont {{Modugno}}},\ }\bibfield  {title} {\bibinfo
  {title} {{Measurement of the mobility edge for 3D Anderson localization}},\
  }\href {https://doi.org/10.1038/nphys3339} {\bibfield  {journal} {\bibinfo
  {journal} {Nat. Phys.}\ }\textbf {\bibinfo {volume} {11}},\ \bibinfo {pages}
  {554} (\bibinfo {year} {2015})}\BibitemShut {NoStop}%
\bibitem [{\citenamefont {{Belinicher}}\ and\ \citenamefont
  {{Sturman}}(1980)}]{Belinicher1980}%
  \BibitemOpen
  \bibfield  {author} {\bibinfo {author} {\bibfnamefont {V.~I.}\ \bibnamefont
  {{Belinicher}}}\ and\ \bibinfo {author} {\bibfnamefont {B.~I.}\ \bibnamefont
  {{Sturman}}},\ }\bibfield  {title} {\bibinfo {title} {{The photogalvanic
  effect in media lacking a center of symmetry}},\ }\href
  {https://doi.org/10.1070/PU1980v023n03ABEH004703} {\bibfield  {journal}
  {\bibinfo  {journal} {Sov. Phys. Usp.}\ }\textbf {\bibinfo {volume} {23}},\
  \bibinfo {pages} {199} (\bibinfo {year} {1980})}\BibitemShut {NoStop}%
\bibitem [{\citenamefont {{von Baltz}}\ and\ \citenamefont
  {{Kraut}}(1981)}]{vonBaltz1981}%
  \BibitemOpen
  \bibfield  {author} {\bibinfo {author} {\bibfnamefont {R.}~\bibnamefont {{von
  Baltz}}}\ and\ \bibinfo {author} {\bibfnamefont {W.}~\bibnamefont
  {{Kraut}}},\ }\bibfield  {title} {\bibinfo {title} {{Theory of the bulk
  photovoltaic effect in pure crystals}},\ }\href
  {https://doi.org/10.1103/PhysRevB.23.5590} {\bibfield  {journal} {\bibinfo
  {journal} {Phys. Rev. B}\ }\textbf {\bibinfo {volume} {23}},\ \bibinfo
  {pages} {5590} (\bibinfo {year} {1981})}\BibitemShut {NoStop}%
\bibitem [{\citenamefont {{Ventura}}\ \emph {et~al.}(2017)\citenamefont
  {{Ventura}}, \citenamefont {{Passos}}, \citenamefont {{Lopes dos Santos}},
  \citenamefont {{Viana Parente Lopes}},\ and\ \citenamefont
  {{Peres}}}]{Ventura2017}%
  \BibitemOpen
  \bibfield  {author} {\bibinfo {author} {\bibfnamefont {G.~B.}\ \bibnamefont
  {{Ventura}}}, \bibinfo {author} {\bibfnamefont {D.~J.}\ \bibnamefont
  {{Passos}}}, \bibinfo {author} {\bibfnamefont {J.~M.~B.}\ \bibnamefont
  {{Lopes dos Santos}}}, \bibinfo {author} {\bibfnamefont {J.~M.}\ \bibnamefont
  {{Viana Parente Lopes}}},\ and\ \bibinfo {author} {\bibfnamefont {N.~M.~R.}\
  \bibnamefont {{Peres}}},\ }\bibfield  {title} {\bibinfo {title} {{Gauge
  covariances and nonlinear optical responses}},\ }\href
  {https://doi.org/10.1103/PhysRevB.96.035431} {\bibfield  {journal} {\bibinfo
  {journal} {Phys. Rev. B}\ }\textbf {\bibinfo {volume} {96}},\ \bibinfo {eid}
  {035431} (\bibinfo {year} {2017})}\BibitemShut {NoStop}%
\bibitem [{\citenamefont {{Parker}}\ \emph {et~al.}(2019)\citenamefont
  {{Parker}}, \citenamefont {{Morimoto}}, \citenamefont {{Orenstein}},\ and\
  \citenamefont {{Moore}}}]{Parker2019}%
  \BibitemOpen
  \bibfield  {author} {\bibinfo {author} {\bibfnamefont {D.~E.}\ \bibnamefont
  {{Parker}}}, \bibinfo {author} {\bibfnamefont {T.}~\bibnamefont
  {{Morimoto}}}, \bibinfo {author} {\bibfnamefont {J.}~\bibnamefont
  {{Orenstein}}},\ and\ \bibinfo {author} {\bibfnamefont {J.~E.}\ \bibnamefont
  {{Moore}}},\ }\bibfield  {title} {\bibinfo {title} {{Diagrammatic approach to
  nonlinear optical response with application to Weyl semimetals}},\ }\href
  {https://doi.org/10.1103/PhysRevB.99.045121} {\bibfield  {journal} {\bibinfo
  {journal} {Phys. Rev. B}\ }\textbf {\bibinfo {volume} {99}},\ \bibinfo {eid}
  {045121} (\bibinfo {year} {2019})}\BibitemShut {NoStop}%
\bibitem [{\citenamefont {{Holder}}\ \emph {et~al.}(2020)\citenamefont
  {{Holder}}, \citenamefont {{Kaplan}},\ and\ \citenamefont
  {{Yan}}}]{Holder2020}%
  \BibitemOpen
  \bibfield  {author} {\bibinfo {author} {\bibfnamefont {T.}~\bibnamefont
  {{Holder}}}, \bibinfo {author} {\bibfnamefont {D.}~\bibnamefont {{Kaplan}}},\
  and\ \bibinfo {author} {\bibfnamefont {B.}~\bibnamefont {{Yan}}},\ }\bibfield
   {title} {\bibinfo {title} {{Consequences of time-reversal-symmetry breaking
  in the light-matter interaction: Berry curvature, quantum metric, and
  diabatic motion}},\ }\href {https://doi.org/10.1103/PhysRevResearch.2.033100}
  {\bibfield  {journal} {\bibinfo  {journal} {Physical Review Research}\
  }\textbf {\bibinfo {volume} {2}},\ \bibinfo {eid} {033100} (\bibinfo {year}
  {2020})}\BibitemShut {NoStop}%
\bibitem [{\citenamefont {Belinicher}\ \emph {et~al.}(1986)\citenamefont
  {Belinicher}, \citenamefont {Ivchenko},\ and\ \citenamefont
  {Pikus}}]{Belinicher1986}%
  \BibitemOpen
  \bibfield  {author} {\bibinfo {author} {\bibfnamefont {V.}~\bibnamefont
  {Belinicher}}, \bibinfo {author} {\bibfnamefont {E.}~\bibnamefont
  {Ivchenko}},\ and\ \bibinfo {author} {\bibfnamefont {G.}~\bibnamefont
  {Pikus}},\ }\bibfield  {title} {\bibinfo {title} {Transient photocurrent in
  gyrotropic crystals},\ }\href@noop {} {\bibfield  {journal} {\bibinfo
  {journal} {Sov. Phys. Semicond.}\ }\textbf {\bibinfo {volume} {20}},\
  \bibinfo {pages} {558} (\bibinfo {year} {1986})}\BibitemShut {NoStop}%
\bibitem [{\citenamefont {{Morimoto}}\ \emph {et~al.}(2016)\citenamefont
  {{Morimoto}}, \citenamefont {{Zhong}}, \citenamefont {{Orenstein}},\ and\
  \citenamefont {{Moore}}}]{Morimoto2016a}%
  \BibitemOpen
  \bibfield  {author} {\bibinfo {author} {\bibfnamefont {T.}~\bibnamefont
  {{Morimoto}}}, \bibinfo {author} {\bibfnamefont {S.}~\bibnamefont {{Zhong}}},
  \bibinfo {author} {\bibfnamefont {J.}~\bibnamefont {{Orenstein}}},\ and\
  \bibinfo {author} {\bibfnamefont {J.~E.}\ \bibnamefont {{Moore}}},\
  }\bibfield  {title} {\bibinfo {title} {{Semiclassical theory of nonlinear
  magneto-optical responses with applications to topological Dirac/Weyl
  semimetals}},\ }\href {https://doi.org/10.1103/PhysRevB.94.245121} {\bibfield
   {journal} {\bibinfo  {journal} {Phys. Rev. B}\ }\textbf {\bibinfo {volume}
  {94}},\ \bibinfo {eid} {245121} (\bibinfo {year} {2016})}\BibitemShut
  {NoStop}%
\bibitem [{\citenamefont {{Tan}}\ \emph {et~al.}(2016)\citenamefont {{Tan}},
  \citenamefont {{Zheng}}, \citenamefont {{Young}}, \citenamefont {{Wang}},
  \citenamefont {{Liu}},\ and\ \citenamefont {{Rappe}}}]{Tan2016}%
  \BibitemOpen
  \bibfield  {author} {\bibinfo {author} {\bibfnamefont {L.~Z.}\ \bibnamefont
  {{Tan}}}, \bibinfo {author} {\bibfnamefont {F.}~\bibnamefont {{Zheng}}},
  \bibinfo {author} {\bibfnamefont {S.~M.}\ \bibnamefont {{Young}}}, \bibinfo
  {author} {\bibfnamefont {F.}~\bibnamefont {{Wang}}}, \bibinfo {author}
  {\bibfnamefont {S.}~\bibnamefont {{Liu}}},\ and\ \bibinfo {author}
  {\bibfnamefont {A.~M.}\ \bibnamefont {{Rappe}}},\ }\bibfield  {title}
  {\bibinfo {title} {{Shift current bulk photovoltaic effect in polar
  materials{\textemdash}hybrid and oxide perovskites and beyond}},\ }\href
  {https://doi.org/10.1038/npjcompumats.2016.26} {\bibfield  {journal}
  {\bibinfo  {journal} {npj Comput. Math.}\ }\textbf {\bibinfo {volume} {2}},\
  \bibinfo {eid} {16026} (\bibinfo {year} {2016})}\BibitemShut {NoStop}%
\bibitem [{\citenamefont {{Cook}}\ \emph {et~al.}(2017)\citenamefont {{Cook}},
  \citenamefont {{M.~Fregoso}}, \citenamefont {{de Juan}}, \citenamefont
  {{Coh}},\ and\ \citenamefont {{Moore}}}]{Cook2017}%
  \BibitemOpen
  \bibfield  {author} {\bibinfo {author} {\bibfnamefont {A.~M.}\ \bibnamefont
  {{Cook}}}, \bibinfo {author} {\bibfnamefont {B.}~\bibnamefont
  {{M.~Fregoso}}}, \bibinfo {author} {\bibfnamefont {F.}~\bibnamefont {{de
  Juan}}}, \bibinfo {author} {\bibfnamefont {S.}~\bibnamefont {{Coh}}},\ and\
  \bibinfo {author} {\bibfnamefont {J.~E.}\ \bibnamefont {{Moore}}},\
  }\bibfield  {title} {\bibinfo {title} {{Design principles for shift current
  photovoltaics}},\ }\href {https://doi.org/10.1038/ncomms14176} {\bibfield
  {journal} {\bibinfo  {journal} {Nat. Comm.}\ }\textbf {\bibinfo {volume}
  {8}},\ \bibinfo {pages} {14176} (\bibinfo {year} {2017})}\BibitemShut
  {NoStop}%
\bibitem [{\citenamefont {{Olbrich}}\ \emph {et~al.}(2009)\citenamefont
  {{Olbrich}}, \citenamefont {{Tarasenko}}, \citenamefont {{Reitmaier}},
  \citenamefont {{Karch}}, \citenamefont {{Plohmann}}, \citenamefont {{Kvon}},\
  and\ \citenamefont {{Ganichev}}}]{Olbrich2009}%
  \BibitemOpen
  \bibfield  {author} {\bibinfo {author} {\bibfnamefont {P.}~\bibnamefont
  {{Olbrich}}}, \bibinfo {author} {\bibfnamefont {S.~A.}\ \bibnamefont
  {{Tarasenko}}}, \bibinfo {author} {\bibfnamefont {C.}~\bibnamefont
  {{Reitmaier}}}, \bibinfo {author} {\bibfnamefont {J.}~\bibnamefont
  {{Karch}}}, \bibinfo {author} {\bibfnamefont {D.}~\bibnamefont {{Plohmann}}},
  \bibinfo {author} {\bibfnamefont {Z.~D.}\ \bibnamefont {{Kvon}}},\ and\
  \bibinfo {author} {\bibfnamefont {S.~D.}\ \bibnamefont {{Ganichev}}},\
  }\bibfield  {title} {\bibinfo {title} {{Observation of the orbital circular
  photogalvanic effect}},\ }\href {https://doi.org/10.1103/PhysRevB.79.121302}
  {\bibfield  {journal} {\bibinfo  {journal} {Phys. Rev. B}\ }\textbf {\bibinfo
  {volume} {79}},\ \bibinfo {eid} {121302} (\bibinfo {year}
  {2009})}\BibitemShut {NoStop}%
\bibitem [{\citenamefont {{Yuan}}\ \emph {et~al.}(2014)\citenamefont {{Yuan}},
  \citenamefont {{Wang}}, \citenamefont {{Lian}}, \citenamefont {{Zhang}},
  \citenamefont {{Fang}}, \citenamefont {{Shen}}, \citenamefont {{Xu}},
  \citenamefont {{Xu}}, \citenamefont {{Zhang}}, \citenamefont {{Hwang}},\ and\
  \citenamefont {{Cui}}}]{Yuan2014}%
  \BibitemOpen
  \bibfield  {author} {\bibinfo {author} {\bibfnamefont {H.}~\bibnamefont
  {{Yuan}}}, \bibinfo {author} {\bibfnamefont {X.}~\bibnamefont {{Wang}}},
  \bibinfo {author} {\bibfnamefont {B.}~\bibnamefont {{Lian}}}, \bibinfo
  {author} {\bibfnamefont {H.}~\bibnamefont {{Zhang}}}, \bibinfo {author}
  {\bibfnamefont {X.}~\bibnamefont {{Fang}}}, \bibinfo {author} {\bibfnamefont
  {B.}~\bibnamefont {{Shen}}}, \bibinfo {author} {\bibfnamefont
  {G.}~\bibnamefont {{Xu}}}, \bibinfo {author} {\bibfnamefont {Y.}~\bibnamefont
  {{Xu}}}, \bibinfo {author} {\bibfnamefont {S.-C.}\ \bibnamefont {{Zhang}}},
  \bibinfo {author} {\bibfnamefont {H.~Y.}\ \bibnamefont {{Hwang}}},\ and\
  \bibinfo {author} {\bibfnamefont {Y.}~\bibnamefont {{Cui}}},\ }\bibfield
  {title} {\bibinfo {title} {{Generation and electric control of
  spin-valley-coupled circular photogalvanic current in WSe$_{2}$}},\ }\href
  {https://doi.org/10.1038/nnano.2014.183} {\bibfield  {journal} {\bibinfo
  {journal} {Nat. Nanotechnol.}\ }\textbf {\bibinfo {volume} {9}},\ \bibinfo
  {pages} {851} (\bibinfo {year} {2014})}\BibitemShut {NoStop}%
\bibitem [{\citenamefont {{McIver}}\ \emph {et~al.}(2012)\citenamefont
  {{McIver}}, \citenamefont {{Hsieh}}, \citenamefont {{Steinberg}},
  \citenamefont {{Jarillo-Herrero}},\ and\ \citenamefont
  {{Gedik}}}]{McIver2012}%
  \BibitemOpen
  \bibfield  {author} {\bibinfo {author} {\bibfnamefont {J.~W.}\ \bibnamefont
  {{McIver}}}, \bibinfo {author} {\bibfnamefont {D.}~\bibnamefont {{Hsieh}}},
  \bibinfo {author} {\bibfnamefont {H.}~\bibnamefont {{Steinberg}}}, \bibinfo
  {author} {\bibfnamefont {P.}~\bibnamefont {{Jarillo-Herrero}}},\ and\
  \bibinfo {author} {\bibfnamefont {N.}~\bibnamefont {{Gedik}}},\ }\bibfield
  {title} {\bibinfo {title} {{Control over topological insulator photocurrents
  with light polarization}},\ }\href {https://doi.org/10.1038/nnano.2011.214}
  {\bibfield  {journal} {\bibinfo  {journal} {Nat. Nanotechnol.}\ }\textbf
  {\bibinfo {volume} {7}},\ \bibinfo {pages} {96} (\bibinfo {year}
  {2012})}\BibitemShut {NoStop}%
\bibitem [{\citenamefont {{Okada}}\ \emph {et~al.}(2016)\citenamefont
  {{Okada}}, \citenamefont {{Ogawa}}, \citenamefont {{Yoshimi}}, \citenamefont
  {{Tsukazaki}}, \citenamefont {{Takahashi}}, \citenamefont {{Kawasaki}},\ and\
  \citenamefont {{Tokura}}}]{Okada2016}%
  \BibitemOpen
  \bibfield  {author} {\bibinfo {author} {\bibfnamefont {K.~N.}\ \bibnamefont
  {{Okada}}}, \bibinfo {author} {\bibfnamefont {N.}~\bibnamefont {{Ogawa}}},
  \bibinfo {author} {\bibfnamefont {R.}~\bibnamefont {{Yoshimi}}}, \bibinfo
  {author} {\bibfnamefont {A.}~\bibnamefont {{Tsukazaki}}}, \bibinfo {author}
  {\bibfnamefont {K.~S.}\ \bibnamefont {{Takahashi}}}, \bibinfo {author}
  {\bibfnamefont {M.}~\bibnamefont {{Kawasaki}}},\ and\ \bibinfo {author}
  {\bibfnamefont {Y.}~\bibnamefont {{Tokura}}},\ }\bibfield  {title} {\bibinfo
  {title} {{Enhanced photogalvanic current in topological insulators via Fermi
  energy tuning}},\ }\href {https://doi.org/10.1103/PhysRevB.93.081403}
  {\bibfield  {journal} {\bibinfo  {journal} {Phys. Rev. B}\ }\textbf {\bibinfo
  {volume} {93}},\ \bibinfo {eid} {081403} (\bibinfo {year}
  {2016})}\BibitemShut {NoStop}%
\bibitem [{\citenamefont {{K{\"o}nig}}\ \emph {et~al.}(2017)\citenamefont
  {{K{\"o}nig}}, \citenamefont {{Xie}}, \citenamefont {{Pesin}},\ and\
  \citenamefont {{Levchenko}}}]{Koenig2017}%
  \BibitemOpen
  \bibfield  {author} {\bibinfo {author} {\bibfnamefont {E.~J.}\ \bibnamefont
  {{K{\"o}nig}}}, \bibinfo {author} {\bibfnamefont {H.~Y.}\ \bibnamefont
  {{Xie}}}, \bibinfo {author} {\bibfnamefont {D.~A.}\ \bibnamefont {{Pesin}}},\
  and\ \bibinfo {author} {\bibfnamefont {A.}~\bibnamefont {{Levchenko}}},\
  }\bibfield  {title} {\bibinfo {title} {{Photogalvanic effect in Weyl
  semimetals}},\ }\href {https://doi.org/10.1103/PhysRevB.96.075123} {\bibfield
   {journal} {\bibinfo  {journal} {Phys. Rev. B}\ }\textbf {\bibinfo {volume}
  {96}},\ \bibinfo {eid} {075123} (\bibinfo {year} {2017})}\BibitemShut
  {NoStop}%
\bibitem [{\citenamefont {{Golub}}\ and\ \citenamefont
  {{Ivchenko}}(2018)}]{Golub2018}%
  \BibitemOpen
  \bibfield  {author} {\bibinfo {author} {\bibfnamefont {L.~E.}\ \bibnamefont
  {{Golub}}}\ and\ \bibinfo {author} {\bibfnamefont {E.~L.}\ \bibnamefont
  {{Ivchenko}}},\ }\bibfield  {title} {\bibinfo {title} {{Circular and
  magnetoinduced photocurrents in Weyl semimetals}},\ }\href
  {https://doi.org/10.1103/PhysRevB.98.075305} {\bibfield  {journal} {\bibinfo
  {journal} {Phys. Rev. B}\ }\textbf {\bibinfo {volume} {98}},\ \bibinfo {eid}
  {075305} (\bibinfo {year} {2018})}\BibitemShut {NoStop}%
\bibitem [{\citenamefont {{de Juan}}\ \emph {et~al.}(2019)\citenamefont {{de
  Juan}}, \citenamefont {{Zhang}}, \citenamefont {{Morimoto}}, \citenamefont
  {{Sun}}, \citenamefont {{Moore}},\ and\ \citenamefont
  {{Grushin}}}]{deJuan2019}%
  \BibitemOpen
  \bibfield  {author} {\bibinfo {author} {\bibfnamefont {F.}~\bibnamefont {{de
  Juan}}}, \bibinfo {author} {\bibfnamefont {Y.}~\bibnamefont {{Zhang}}},
  \bibinfo {author} {\bibfnamefont {T.}~\bibnamefont {{Morimoto}}}, \bibinfo
  {author} {\bibfnamefont {Y.}~\bibnamefont {{Sun}}}, \bibinfo {author}
  {\bibfnamefont {J.~E.}\ \bibnamefont {{Moore}}},\ and\ \bibinfo {author}
  {\bibfnamefont {A.~G.}\ \bibnamefont {{Grushin}}},\ }\bibfield  {title}
  {\bibinfo {title} {{Difference frequency generation in topological
  semimetals}},\ }\href@noop {} {\bibfield  {journal} {\bibinfo  {journal}
  {arXiv}\ ,\ \bibinfo {eid} {arXiv:1907.02537}} (\bibinfo {year}
  {2019})}\BibitemShut {NoStop}%
\bibitem [{\citenamefont {{Bhalla}}\ \emph {et~al.}(2020)\citenamefont
  {{Bhalla}}, \citenamefont {{MacDonald}},\ and\ \citenamefont
  {{Culcer}}}]{Bhalla2020}%
  \BibitemOpen
  \bibfield  {author} {\bibinfo {author} {\bibfnamefont {P.}~\bibnamefont
  {{Bhalla}}}, \bibinfo {author} {\bibfnamefont {A.~H.}\ \bibnamefont
  {{MacDonald}}},\ and\ \bibinfo {author} {\bibfnamefont {D.}~\bibnamefont
  {{Culcer}}},\ }\bibfield  {title} {\bibinfo {title} {{Resonant Photovoltaic
  Effect in Doped Magnetic Semiconductors}},\ }\href
  {https://doi.org/10.1103/PhysRevLett.124.087402} {\bibfield  {journal}
  {\bibinfo  {journal} {Phys. Rev. Lett.}\ }\textbf {\bibinfo {volume} {124}},\
  \bibinfo {eid} {087402} (\bibinfo {year} {2020})}\BibitemShut {NoStop}%
\bibitem [{Note1()}]{Note1}%
  \BibitemOpen
  \bibinfo {note} {Unless stated otherwise, we do not specify whether these
  lifetimes are quasiparticle lifetimes or momentum lifetimes. Such distinction
  can be easily made from the diagrammatics~\cite {Holder2020}. Also note that
  the bulk-photovoltaic effect does not rely on carriers in the conduction
  band, which is why the interband scattering rate (recombination rate) does
  not enter.}\BibitemShut {Stop}%
\bibitem [{\citenamefont {{Young}}\ and\ \citenamefont
  {{Rappe}}(2012)}]{Young2012}%
  \BibitemOpen
  \bibfield  {author} {\bibinfo {author} {\bibfnamefont {S.~M.}\ \bibnamefont
  {{Young}}}\ and\ \bibinfo {author} {\bibfnamefont {A.~M.}\ \bibnamefont
  {{Rappe}}},\ }\bibfield  {title} {\bibinfo {title} {{First Principles
  Calculation of the Shift Current Photovoltaic Effect in Ferroelectrics}},\
  }\href {https://doi.org/10.1103/PhysRevLett.109.116601} {\bibfield  {journal}
  {\bibinfo  {journal} {Phys. Rev. Lett.}\ }\textbf {\bibinfo {volume} {109}},\
  \bibinfo {eid} {116601} (\bibinfo {year} {2012})}\BibitemShut {NoStop}%
\bibitem [{Note2()}]{Note2}%
  \BibitemOpen
  \bibinfo {note} {See supplemental material for details of this
  derivation.}\BibitemShut {Stop}%
\bibitem [{Note3()}]{Note3}%
  \BibitemOpen
  \bibinfo {note} {In materials where the transport time and quasiparticle
  lifetime are dissimilar, it is necessary to also account for possible vertex
  corrections. This does not change the main observation put forth here that
  only the adiabatic limit leads to a vanishing sub-gap response.}\BibitemShut
  {Stop}%
\bibitem [{\citenamefont {{Kane}}\ and\ \citenamefont
  {{Mele}}(2005)}]{Kane2005}%
  \BibitemOpen
  \bibfield  {author} {\bibinfo {author} {\bibfnamefont {C.~L.}\ \bibnamefont
  {{Kane}}}\ and\ \bibinfo {author} {\bibfnamefont {E.~J.}\ \bibnamefont
  {{Mele}}},\ }\bibfield  {title} {\bibinfo {title} {{Quantum Spin Hall Effect
  in Graphene}},\ }\href {https://doi.org/10.1103/PhysRevLett.95.226801}
  {\bibfield  {journal} {\bibinfo  {journal} {Phys. Rev. Lett.}\ }\textbf
  {\bibinfo {volume} {95}},\ \bibinfo {eid} {226801} (\bibinfo {year}
  {2005})}\BibitemShut {NoStop}%
\bibitem [{\citenamefont {{Bernevig}}\ and\ \citenamefont
  {{Zhang}}(2006)}]{Bernevig2006}%
  \BibitemOpen
  \bibfield  {author} {\bibinfo {author} {\bibfnamefont {B.~A.}\ \bibnamefont
  {{Bernevig}}}\ and\ \bibinfo {author} {\bibfnamefont {S.-C.}\ \bibnamefont
  {{Zhang}}},\ }\bibfield  {title} {\bibinfo {title} {{Quantum Spin Hall
  Effect}},\ }\href {https://doi.org/10.1103/PhysRevLett.96.106802} {\bibfield
  {journal} {\bibinfo  {journal} {Phys. Rev. Lett.}\ }\textbf {\bibinfo
  {volume} {96}},\ \bibinfo {eid} {106802} (\bibinfo {year}
  {2006})}\BibitemShut {NoStop}%
\bibitem [{\citenamefont {{Licciardello}}\ and\ \citenamefont
  {{Thouless}}(1975)}]{Licciardello1975}%
  \BibitemOpen
  \bibfield  {author} {\bibinfo {author} {\bibfnamefont {D.~C.}\ \bibnamefont
  {{Licciardello}}}\ and\ \bibinfo {author} {\bibfnamefont {D.~J.}\
  \bibnamefont {{Thouless}}},\ }\bibfield  {title} {\bibinfo {title}
  {{Conductivity and mobility edges for two-dimensional disordered systems}},\
  }\href {https://doi.org/10.1088/0022-3719/8/24/009} {\bibfield  {journal}
  {\bibinfo  {journal} {J. Phys. C Solid State Phys.}\ }\textbf {\bibinfo
  {volume} {8}},\ \bibinfo {pages} {4157} (\bibinfo {year} {1975})}\BibitemShut
  {NoStop}%
\bibitem [{\citenamefont {{Passos}}\ \emph {et~al.}(2018)\citenamefont
  {{Passos}}, \citenamefont {{Ventura}}, \citenamefont {{Lopes}}, \citenamefont
  {{Santos}},\ and\ \citenamefont {{Peres}}}]{Passos2018}%
  \BibitemOpen
  \bibfield  {author} {\bibinfo {author} {\bibfnamefont {D.~J.}\ \bibnamefont
  {{Passos}}}, \bibinfo {author} {\bibfnamefont {G.~B.}\ \bibnamefont
  {{Ventura}}}, \bibinfo {author} {\bibfnamefont {J.~M. V.~P.}\ \bibnamefont
  {{Lopes}}}, \bibinfo {author} {\bibfnamefont {J.~M.~B. L.~d.}\ \bibnamefont
  {{Santos}}},\ and\ \bibinfo {author} {\bibfnamefont {N.~M.~R.}\ \bibnamefont
  {{Peres}}},\ }\bibfield  {title} {\bibinfo {title} {{Nonlinear optical
  responses of crystalline systems: Results from a velocity gauge analysis}},\
  }\href {https://doi.org/10.1103/PhysRevB.97.235446} {\bibfield  {journal}
  {\bibinfo  {journal} {Phys. Rev. B}\ }\textbf {\bibinfo {volume} {97}},\
  \bibinfo {eid} {235446} (\bibinfo {year} {2018})}\BibitemShut {NoStop}%
\bibitem [{\citenamefont {{Piraud}}\ \emph {et~al.}(2013)\citenamefont
  {{Piraud}}, \citenamefont {{Pezz{\'e}}},\ and\ \citenamefont
  {{Sanchez-Palencia}}}]{Piraud2013}%
  \BibitemOpen
  \bibfield  {author} {\bibinfo {author} {\bibfnamefont {M.}~\bibnamefont
  {{Piraud}}}, \bibinfo {author} {\bibfnamefont {L.}~\bibnamefont
  {{Pezz{\'e}}}},\ and\ \bibinfo {author} {\bibfnamefont {L.}~\bibnamefont
  {{Sanchez-Palencia}}},\ }\bibfield  {title} {\bibinfo {title} {{Quantum
  transport of atomic matter waves in anisotropic two-dimensional and
  three-dimensional disorder}},\ }\href
  {https://doi.org/10.1088/1367-2630/15/7/075007} {\bibfield  {journal}
  {\bibinfo  {journal} {New J. Phys.}\ }\textbf {\bibinfo {volume} {15}},\
  \bibinfo {eid} {075007} (\bibinfo {year} {2013})}\BibitemShut {NoStop}%
\bibitem [{\citenamefont {{M{\"u}ller}}\ \emph {et~al.}(2016)\citenamefont
  {{M{\"u}ller}}, \citenamefont {{Delande}},\ and\ \citenamefont
  {{Shapiro}}}]{Mueller2016}%
  \BibitemOpen
  \bibfield  {author} {\bibinfo {author} {\bibfnamefont {C.~A.}\ \bibnamefont
  {{M{\"u}ller}}}, \bibinfo {author} {\bibfnamefont {D.}~\bibnamefont
  {{Delande}}},\ and\ \bibinfo {author} {\bibfnamefont {B.}~\bibnamefont
  {{Shapiro}}},\ }\bibfield  {title} {\bibinfo {title} {{Critical dynamics at
  the Anderson localization mobility edge}},\ }\href
  {https://doi.org/10.1103/PhysRevA.94.033615} {\bibfield  {journal} {\bibinfo
  {journal} {Phys. Rev. A}\ }\textbf {\bibinfo {volume} {94}},\ \bibinfo {eid}
  {033615} (\bibinfo {year} {2016})}\BibitemShut {NoStop}%
\bibitem [{\citenamefont {Mahan}(1990)}]{Mahan1990}%
  \BibitemOpen
  \bibfield  {author} {\bibinfo {author} {\bibfnamefont {G.}~\bibnamefont
  {Mahan}},\ }\href@noop {} {\emph {\bibinfo {title} {Many-Particle Physics}}}\
  (\bibinfo  {publisher} {Springer},\ \bibinfo {year} {1990})\BibitemShut
  {NoStop}%
\bibitem [{\citenamefont {{Ganichev}}\ \emph {et~al.}(2002)\citenamefont
  {{Ganichev}}, \citenamefont {{Ivchenko}}, \citenamefont {{Bel'kov}},
  \citenamefont {{Tarasenko}}, \citenamefont {{Sollinger}}, \citenamefont
  {{Weiss}}, \citenamefont {{Wegscheider}},\ and\ \citenamefont
  {{Prettl}}}]{Ganichev2002}%
  \BibitemOpen
  \bibfield  {author} {\bibinfo {author} {\bibfnamefont {S.~D.}\ \bibnamefont
  {{Ganichev}}}, \bibinfo {author} {\bibfnamefont {E.~L.}\ \bibnamefont
  {{Ivchenko}}}, \bibinfo {author} {\bibfnamefont {V.~V.}\ \bibnamefont
  {{Bel'kov}}}, \bibinfo {author} {\bibfnamefont {S.~A.}\ \bibnamefont
  {{Tarasenko}}}, \bibinfo {author} {\bibfnamefont {M.}~\bibnamefont
  {{Sollinger}}}, \bibinfo {author} {\bibfnamefont {D.}~\bibnamefont
  {{Weiss}}}, \bibinfo {author} {\bibfnamefont {W.}~\bibnamefont
  {{Wegscheider}}},\ and\ \bibinfo {author} {\bibfnamefont {W.}~\bibnamefont
  {{Prettl}}},\ }\bibfield  {title} {\bibinfo {title} {{Spin-galvanic
  effect}},\ }\href {https://doi.org/10.1038/417153a} {\bibfield  {journal}
  {\bibinfo  {journal} {Nature}\ }\textbf {\bibinfo {volume} {417}},\ \bibinfo
  {pages} {153} (\bibinfo {year} {2002})}\BibitemShut {NoStop}%
\bibitem [{\citenamefont {{Ganichev}}\ \emph {et~al.}(2006)\citenamefont
  {{Ganichev}}, \citenamefont {{Bel'Kov}}, \citenamefont {{Tarasenko}},
  \citenamefont {{Danilov}}, \citenamefont {{Giglberger}}, \citenamefont
  {{Hoffmann}}, \citenamefont {{Ivchenko}}, \citenamefont {{Weiss}},
  \citenamefont {{Wegscheider}}, \citenamefont {{Gerl}}, \citenamefont
  {{Schuh}}, \citenamefont {{Stahl}}, \citenamefont {{de Boeck}}, \citenamefont
  {{Borghs}},\ and\ \citenamefont {{Prettl}}}]{Ganichev2006}%
  \BibitemOpen
  \bibfield  {author} {\bibinfo {author} {\bibfnamefont {S.~D.}\ \bibnamefont
  {{Ganichev}}}, \bibinfo {author} {\bibfnamefont {V.~V.}\ \bibnamefont
  {{Bel'Kov}}}, \bibinfo {author} {\bibfnamefont {S.~A.}\ \bibnamefont
  {{Tarasenko}}}, \bibinfo {author} {\bibfnamefont {S.~N.}\ \bibnamefont
  {{Danilov}}}, \bibinfo {author} {\bibfnamefont {S.}~\bibnamefont
  {{Giglberger}}}, \bibinfo {author} {\bibfnamefont {C.}~\bibnamefont
  {{Hoffmann}}}, \bibinfo {author} {\bibfnamefont {E.~L.}\ \bibnamefont
  {{Ivchenko}}}, \bibinfo {author} {\bibfnamefont {D.}~\bibnamefont {{Weiss}}},
  \bibinfo {author} {\bibfnamefont {W.}~\bibnamefont {{Wegscheider}}}, \bibinfo
  {author} {\bibfnamefont {C.}~\bibnamefont {{Gerl}}}, \bibinfo {author}
  {\bibfnamefont {D.}~\bibnamefont {{Schuh}}}, \bibinfo {author} {\bibfnamefont
  {J.}~\bibnamefont {{Stahl}}}, \bibinfo {author} {\bibfnamefont
  {J.}~\bibnamefont {{de Boeck}}}, \bibinfo {author} {\bibfnamefont
  {G.}~\bibnamefont {{Borghs}}},\ and\ \bibinfo {author} {\bibfnamefont
  {W.}~\bibnamefont {{Prettl}}},\ }\bibfield  {title} {\bibinfo {title}
  {{Zero-bias spin separation}},\ }\href {https://doi.org/10.1038/nphys390}
  {\bibfield  {journal} {\bibinfo  {journal} {Nat. Phys.}\ }\textbf {\bibinfo
  {volume} {2}},\ \bibinfo {pages} {609} (\bibinfo {year} {2006})}\BibitemShut
  {NoStop}%
\bibitem [{\citenamefont {{Zhang}}\ \emph {et~al.}(2019)\citenamefont
  {{Zhang}}, \citenamefont {{Holder}}, \citenamefont {{Ishizuka}},
  \citenamefont {{de Juan}}, \citenamefont {{Nagaosa}}, \citenamefont
  {{Felser}},\ and\ \citenamefont {{Yan}}}]{Zhang2019}%
  \BibitemOpen
  \bibfield  {author} {\bibinfo {author} {\bibfnamefont {Y.}~\bibnamefont
  {{Zhang}}}, \bibinfo {author} {\bibfnamefont {T.}~\bibnamefont {{Holder}}},
  \bibinfo {author} {\bibfnamefont {H.}~\bibnamefont {{Ishizuka}}}, \bibinfo
  {author} {\bibfnamefont {F.}~\bibnamefont {{de Juan}}}, \bibinfo {author}
  {\bibfnamefont {N.}~\bibnamefont {{Nagaosa}}}, \bibinfo {author}
  {\bibfnamefont {C.}~\bibnamefont {{Felser}}},\ and\ \bibinfo {author}
  {\bibfnamefont {B.}~\bibnamefont {{Yan}}},\ }\bibfield  {title} {\bibinfo
  {title} {{Switchable magnetic bulk photovoltaic effect in the two-dimensional
  magnet CrI$_{3}$}},\ }\href {https://doi.org/10.1038/s41467-019-11832-3}
  {\bibfield  {journal} {\bibinfo  {journal} {Nat. Comm.}\ }\textbf {\bibinfo
  {volume} {10}},\ \bibinfo {eid} {3783} (\bibinfo {year} {2019})}\BibitemShut
  {NoStop}%
\bibitem [{\citenamefont {{Sun}}\ \emph {et~al.}(2019)\citenamefont {{Sun}},
  \citenamefont {{Yi}}, \citenamefont {{Song}}, \citenamefont {{Clark}},
  \citenamefont {{Huang}}, \citenamefont {{Shan}}, \citenamefont {{Wu}},
  \citenamefont {{Huang}}, \citenamefont {{Gao}}, \citenamefont {{Chen}},
  \citenamefont {{McGuire}}, \citenamefont {{Cao}}, \citenamefont {{Xiao}},
  \citenamefont {{Liu}}, \citenamefont {{Yao}}, \citenamefont {{Xu}},\ and\
  \citenamefont {{Wu}}}]{Sun2019}%
  \BibitemOpen
  \bibfield  {author} {\bibinfo {author} {\bibfnamefont {Z.}~\bibnamefont
  {{Sun}}}, \bibinfo {author} {\bibfnamefont {Y.}~\bibnamefont {{Yi}}},
  \bibinfo {author} {\bibfnamefont {T.}~\bibnamefont {{Song}}}, \bibinfo
  {author} {\bibfnamefont {G.}~\bibnamefont {{Clark}}}, \bibinfo {author}
  {\bibfnamefont {B.}~\bibnamefont {{Huang}}}, \bibinfo {author} {\bibfnamefont
  {Y.}~\bibnamefont {{Shan}}}, \bibinfo {author} {\bibfnamefont
  {S.}~\bibnamefont {{Wu}}}, \bibinfo {author} {\bibfnamefont {D.}~\bibnamefont
  {{Huang}}}, \bibinfo {author} {\bibfnamefont {C.}~\bibnamefont {{Gao}}},
  \bibinfo {author} {\bibfnamefont {Z.}~\bibnamefont {{Chen}}}, \bibinfo
  {author} {\bibfnamefont {M.}~\bibnamefont {{McGuire}}}, \bibinfo {author}
  {\bibfnamefont {T.}~\bibnamefont {{Cao}}}, \bibinfo {author} {\bibfnamefont
  {D.}~\bibnamefont {{Xiao}}}, \bibinfo {author} {\bibfnamefont {W.-T.}\
  \bibnamefont {{Liu}}}, \bibinfo {author} {\bibfnamefont {W.}~\bibnamefont
  {{Yao}}}, \bibinfo {author} {\bibfnamefont {X.}~\bibnamefont {{Xu}}},\ and\
  \bibinfo {author} {\bibfnamefont {S.}~\bibnamefont {{Wu}}},\ }\bibfield
  {title} {\bibinfo {title} {{Giant nonreciprocal second-harmonic generation
  from antiferromagnetic bilayer CrI$_{3}$}},\ }\href
  {https://doi.org/10.1038/s41586-019-1445-3} {\bibfield  {journal} {\bibinfo
  {journal} {Nature}\ }\textbf {\bibinfo {volume} {572}},\ \bibinfo {pages}
  {497} (\bibinfo {year} {2019})}\BibitemShut {NoStop}%
\bibitem [{\citenamefont {Li}\ \emph {et~al.}(2012)\citenamefont {Li},
  \citenamefont {Cao}, \citenamefont {Niu}, \citenamefont {Shi},\ and\
  \citenamefont {Feng}}]{Li2012}%
  \BibitemOpen
  \bibfield  {author} {\bibinfo {author} {\bibfnamefont {X.}~\bibnamefont
  {Li}}, \bibinfo {author} {\bibfnamefont {T.}~\bibnamefont {Cao}}, \bibinfo
  {author} {\bibfnamefont {Q.}~\bibnamefont {Niu}}, \bibinfo {author}
  {\bibfnamefont {J.}~\bibnamefont {Shi}},\ and\ \bibinfo {author}
  {\bibfnamefont {J.}~\bibnamefont {Feng}},\ }\bibfield  {title} {\bibinfo
  {title} {{Coupling the valley degree of freedom to antiferromagnetic
  order}},\ }\href {https://doi.org/10.1073/pnas.1219420110} {\bibfield
  {journal} {\bibinfo  {journal} {PNAS}\ }\textbf {\bibinfo {volume} {110}},\
  \bibinfo {pages} {3738} (\bibinfo {year} {2012})}\BibitemShut {NoStop}%
\bibitem [{\citenamefont {{Koitzsch}}\ \emph {et~al.}(2016)\citenamefont
  {{Koitzsch}}, \citenamefont {{Habenicht}}, \citenamefont {{M{\"u}ller}},
  \citenamefont {{Knupfer}}, \citenamefont {{B{\"u}chner}}, \citenamefont
  {{Kandpal}}, \citenamefont {{van den Brink}}, \citenamefont {{Nowak}},
  \citenamefont {{Isaeva}},\ and\ \citenamefont {{Doert}}}]{Koitzsch2016}%
  \BibitemOpen
  \bibfield  {author} {\bibinfo {author} {\bibfnamefont {A.}~\bibnamefont
  {{Koitzsch}}}, \bibinfo {author} {\bibfnamefont {C.}~\bibnamefont
  {{Habenicht}}}, \bibinfo {author} {\bibfnamefont {E.}~\bibnamefont
  {{M{\"u}ller}}}, \bibinfo {author} {\bibfnamefont {M.}~\bibnamefont
  {{Knupfer}}}, \bibinfo {author} {\bibfnamefont {B.}~\bibnamefont
  {{B{\"u}chner}}}, \bibinfo {author} {\bibfnamefont {H.~C.}\ \bibnamefont
  {{Kandpal}}}, \bibinfo {author} {\bibfnamefont {J.}~\bibnamefont {{van den
  Brink}}}, \bibinfo {author} {\bibfnamefont {D.}~\bibnamefont {{Nowak}}},
  \bibinfo {author} {\bibfnamefont {A.}~\bibnamefont {{Isaeva}}},\ and\
  \bibinfo {author} {\bibfnamefont {T.}~\bibnamefont {{Doert}}},\ }\bibfield
  {title} {\bibinfo {title} {{J$_{eff}$ Description of the Honeycomb Mott
  Insulator {\ensuremath{\alpha}} -RuCl$_{3}$}},\ }\href
  {https://doi.org/10.1103/PhysRevLett.117.126403} {\bibfield  {journal}
  {\bibinfo  {journal} {Phys. Rev. Lett.}\ }\textbf {\bibinfo {volume} {117}},\
  \bibinfo {eid} {126403} (\bibinfo {year} {2016})}\BibitemShut {NoStop}%
\bibitem [{\citenamefont {{Zhou}}\ \emph {et~al.}(2016)\citenamefont {{Zhou}},
  \citenamefont {{Li}}, \citenamefont {{Waugh}}, \citenamefont {{Parham}},
  \citenamefont {{Kim}}, \citenamefont {{Sears}}, \citenamefont {{Gomes}},
  \citenamefont {{Kee}}, \citenamefont {{Kim}},\ and\ \citenamefont
  {{Dessau}}}]{Zhou2016}%
  \BibitemOpen
  \bibfield  {author} {\bibinfo {author} {\bibfnamefont {X.}~\bibnamefont
  {{Zhou}}}, \bibinfo {author} {\bibfnamefont {H.}~\bibnamefont {{Li}}},
  \bibinfo {author} {\bibfnamefont {J.~A.}\ \bibnamefont {{Waugh}}}, \bibinfo
  {author} {\bibfnamefont {S.}~\bibnamefont {{Parham}}}, \bibinfo {author}
  {\bibfnamefont {H.-S.}\ \bibnamefont {{Kim}}}, \bibinfo {author}
  {\bibfnamefont {J.~A.}\ \bibnamefont {{Sears}}}, \bibinfo {author}
  {\bibfnamefont {A.}~\bibnamefont {{Gomes}}}, \bibinfo {author} {\bibfnamefont
  {H.-Y.}\ \bibnamefont {{Kee}}}, \bibinfo {author} {\bibfnamefont {Y.-J.}\
  \bibnamefont {{Kim}}},\ and\ \bibinfo {author} {\bibfnamefont {D.~S.}\
  \bibnamefont {{Dessau}}},\ }\bibfield  {title} {\bibinfo {title}
  {{Angle-resolved photoemission study of the Kitaev candidate
  {\ensuremath{\alpha}} -RuCl$_{3}$}},\ }\href
  {https://doi.org/10.1103/PhysRevB.94.161106} {\bibfield  {journal} {\bibinfo
  {journal} {Phys. Rev. B}\ }\textbf {\bibinfo {volume} {94}},\ \bibinfo {eid}
  {161106} (\bibinfo {year} {2016})}\BibitemShut {NoStop}%
\end{thebibliography}
\end{document}